\documentclass[traditabstract]{aa}
\usepackage{graphicx}
\usepackage{txfonts}
\begin{document}
   \title{FIR colours of nearby late-type galaxies in the $Herschel$ Reference Survey \thanks{$Herschel$ 
   is an ESA space observatory with science instruments provided by European-led Principal Investigator 
   consortia and with important participation from NASA.}}

   \subtitle{}

  \author{A. Boselli\inst{1}
          ,
	  L. Ciesla\inst{1}
	  ,
	  L. Cortese\inst{2}
	  , 
	  V. Buat\inst{1}
	  ,
	  M. Boquien\inst{1}
	  ,
	  G. J. Bendo\inst{3}
	  ,
	  S. Boissier\inst{1}
	  ,
	  S. Eales\inst{4}
	  ,
	  G. Gavazzi\inst{5}
	  ,
	  T.M. Hughes\inst{6}
	  ,
	  M. Pohlen\inst{4}
	  ,
	  M.W.L. Smith\inst{4}
	  ,
	  M. Baes\inst{7}
	  ,
	  S. Bianchi\inst{8}
	  ,
	  D.L. Clements\inst{9}
	  ,
	  A. Cooray\inst{10}
	  ,
	  J. Davies\inst{4}
	  ,
	  W. Gear\inst{4}
	  ,
	  S. Madden\inst{11}
	  ,
	  L. Magrini\inst{8}
	  ,
	  P. Panuzzo\inst{11}
	  ,
	  A. Remy\inst{11}
	  ,
	  L. Spinoglio\inst{12}
	  ,
	  S. Zibetti\inst{13}
        }

\institute{Laboratoire d'Astrophysique de Marseille - LAM, Universit\'e Aix-Marseille \& CNRS, UMR7326, 38 rue F. Joliot-Curie, 13388 Marseille Cedex 13, France
              \email{Alessandro.Boselli@oamp.fr, Laure.Ciesla@oamp.fr, Veronique.Buat@oamp.fr, Mederic.Boquien@oamp.fr, Samuel.Boissier@oamp.fr}
         \and 
	 European Southern Observatory, Karl-Schwarzschild Str. 2, 85748 Garching bei Muenchen, Germany
                \email{lcortese@eso.org}
	\and
	UK ALMA Regional Centre Node, Jodrell Bank Centre for Astrophysics, School of Physics and Astronomy, University of Manchester, Oxford Road, Manchester M139PL,United Kingdom
		\email{george.bendo@manchester.ac.uk}
	\and 
	School of Physics and Astronomy, Cardiff University, Queens Buildings The Parade, Cardiff CF24 3AA, UK
             \email{Steve.Eales@astro.cf.ac.uk, Michael.Pohlen@astro.cf.ac.uk, Matthew.Smith@astro.cf.ac.uk, Jonathan.Davies@astro.cf.ac.uk, Walter.Gear@astro.cf.ac.uk}
	 \and
	 Universita' di Milano-Bicocca, piazza della Scienza 3, 20100, Milano, Italy
	 	\email{giuseppe.gavazzi@mib.infn.it} 
	 \and
	 Kavli Institute for Astronomy \& Astrophysics, Peking University, Beijing 100871, P.R. China
                \email{tmhughes@pku.edu.cn}
	\and
	Sterrenkundig Observatorium, Universiteit Gent, Krijgslaan 281 S9, B-9000 Gent, Belgium
                \email{maarten.baes@ugent.be}
        \and
	INAF-Osservatorio Astrofisico di Arcetri, Largo Enrico Fermi 5, 50125 Firenze, Italy
		\email{sbianchi@arcetri.astro.it,laura@arcetri.astro.it}
	\and
	Astrophysics Group, Imperial College, Blackett Laboratory, Prince Consort Road, London SW7 2AZ, UK
             \email{d.clements@imperial.ac.uk}
	\and
	Department of Physics and Astronomy, University of California, Irvine, CA 92697, USA ; California Institute of Technology, 1200 E. California Blvd., Pasadena, CA 91125, USA 
		\email{acooray@uci.edu}	
	\and
	CEA/DSM/IRFU/Service d'Astrophysique, CEA, Saclay, Orme des Merisiers, Batiment 709, F-91191 Gif-sur-Yvette, France
			\email{pasquale.panuzzo@cea.fr, aurelie.remy@cea.fr}
	\and
	INAF-Istituto di Astrofisica Spaziale e Fisica Cosmica, via Fosso del Cavaliere 100, I-00133, Roma, Italy 
			\email{luigispinoglio@gmail.com}
	\and
	Dark Cosmology Centre, Niels Bohr Institute - University of Copenhagen, Juliane Maries Vej 30, DK-2100 Copenhagen, Denmark
		\email{zibetti@dark-cosmology.dk}
}

   \date{}

 
  \abstract
{We study the far infrared (60-500 $\mu$m) colours of late-type galaxies in the $Herschel$ Reference Survey, a K-band selected, volume limited
sample of nearby galaxies. The far infrared colours are correlated with each other, with tighter correlations for the indices that are closer in wavelength.
We also compare the different colour indices to various tracers of the physical properties of the target galaxies, such as the surface brightness of the ionising and non-ionising stellar
radiation, the dust attenuation and the metallicity. 
The emission properties of the cold dust dominating the far infrared spectral domain are 
regulated by the properties of the interstellar radiation field. Consistent with that observed in nearby, resolved galaxies, our analysis shows that
the ionising and the non-ionising stellar radiation, including that emitted by the most evolved, cold stars, both contribute to the heating of the cold dust component.
This work also shows that metallicity is another key parameter characterising the cold dust emission of 
normal, late-type galaxies.\\
A single modified black body with a grain
emissivity index $\beta$=1.5 better fits the observed SPIRE flux density ratios $S250/S350$ vs. $S350/S500$ than $\beta$=2, although values of $\beta$ $\simeq$ 2 are possible in metal rich,
high surface brightness galaxies. Values of $\beta$ $\lesssim$ 1.5 better represent metal poor, low surface brightness objects.
This observational evidence provides strong constraints for dust emission models of normal, late type galaxies.\\
}
   {}
   {}
   {}
   {}

	\authorrunning{Boselli et al.}
   \maketitle
%

\section{Introduction}

The emission of late-type galaxies in the infrared domain (5 $\leq$ $\lambda$ $\leq$ 1000 $\mu$m) is dominated by the dust component of the interstellar medium
heated mainly by the stellar radiation field. Young stars are the dominant heating sources in star forming objects, while evolved stars can have a major contribution in quiescent systems.
Although not dominant in mass, dust plays a major role in the equilibrium of the interstellar medium (Sauvage \& Thuan 1992). 
Formed by the aggregation of metals produced by massive and asymptotic giant branch (AGB) stars at the latest stages of their evolution (Valiante et al. 2009)
and injected into the interstellar medium through stellar winds and supernovae explosions, 
dust grains are important coolers of the gaseous phase (Bakes \& Tielens 1994). 
They also act as catalysts in the process responsible for the formation of the molecular hydrogen within molecular clouds,
and are thus of paramount importance in the process of star formation (Hollenbach \& Salpeter 1971). Heated 
by the stellar radiation produced mainly by newly formed stars, the dust emission is often used as a powerful tracer of the star formation 
activity of galaxies, in particular in highly extinguished objects such as dusty starbursts and ultraluminous infrared galaxies (e.g. Kennicutt 1998).\\

Dust is composed of a mixture of carbonaceous and amorphous silicate grains of different size and
composition (see for a review Draine 2003). 
Big grains, of size $a$ $>$ 200 \AA, composed of carbon (graphite or amorphous) and silicates,
are in thermal equilibrium with the UV and the optical radiation field and dominate the IR emission for $\lambda$ $>$ 60 $\mu$m. While the energy output 
of galaxies at wavelengths $\lambda$ $\lesssim$ 200 $\mu$m is dominated by the warm dust, at longer wavelengths 
($\lambda$ $\gtrsim$ 200 $\mu$m) is due to the cold dust emission dominating in mass (e.g. Whittet 1992).\\

The IRAS (Neugebauer et al. 1984), ISO (Kessler et al. 1996), Spitzer (Werner et al. 2004) and more recently AKARI (Murakami et al. 2007) 
space missions allowed us to gather data for hundreds of thousands of extragalactic sources in the spectral domain
$\lambda$ $\leq$ 170 $\mu$m. Sub-millimetric ground based facilities such as SCUBA on the JCMT telescope allowed the observation of the coldest dust at $\lambda$ $\simeq$ 850 $\mu$m. 
The $Herschel$ Space Observatory (Pilbratt et al. 2010), launched in May 2009, has been designed to extend previous imaging and spectroscopic 
infrared observations in the spectral range 55-672 $\mu$m. 
Within the SPIRE and PACS guaranteed time consortia, we have started several observing programs using the three $Herschel$ instruments (SPIRE, Griffin et al. 2010;
PACS, Poglitsch et al. 2010 and HIFI, de Graauw et al. 2010) with the aim of studying the dust emission properties of different types of galaxies. 
The targets of these programmes have been selected to sample the largest parameter space in morphological type, luminosity, stellar and nuclear activity, 
metallicity and environment. One of these projects, the $Herschel$ Reference Survey (HRS), has been designed to cover a complete, 
volume limited, statistically significant sample of nearby galaxies spanning the whole Hubble sequence of morphological type. The aim of this project is to derive the 
mean statistical properties such as the far infrared luminosity distribution, the scaling relations, the colours, and the spectral energy distributions 
of a representative sample of galaxies in the nearby universe (Boselli et al. 2010a). Its completeness makes the HRS the ideal sample for such a purpose.

The present paper studies the broad band far infrared colours 
of late-type systems from the HRS and their relationships with different tracers of the physical properties of the observed galaxies. SPIRE data obtained at
250, 350 and 500 $\mu$m are combined with IRAS 60 and 100 $\mu$m data and with other multifrequency data 
to trace the empirical properties of the dust emission of galaxies in this spectral range. The main aim of this work is that of identifying, through a
multifrequency statistical analysis, the physical parameters driving the dust emission properties in galaxies. This approach is useful in the study of the interstellar
medium for understanding the intrinsic properties of the emitting dust in different types of galaxies. This analysis is also crucial to understand 
whether the standard assumptions taken in the interpretation of cosmological data are justified. The main result of this work is that 
the dust emission properties of galaxies are not universal but strongly depend on the metallicity and the intensity and the hardness of the interstellar radiation field.
Understanding and quantifying these effects in the local universe is thus important for high redshift studies given the strong evolution of the physical properties of galaxies 
with cosmic time. Furthermore, observed far infrared colour indices can be easily compared to those of high redshift sources or of templates generally used 
to characterise the spectral energy distribution of galaxies at any redshift and thus to directly check their validity (Boselli et al. 2010b).\\ 

We limit the analysis to late-type systems, while those of early-types (ellipticals, lenticulars) will be addressed in another communication (Smith et al. 2012).
This choice is dictated by the fact that both the dust properties and the nature of the heating sources in early-type systems are generally significantly
different from those of spiral galaxies.

The present work extends the preliminary study of Boselli et al. (2010b) to the complete HRS sample. At the same time, it widens the analysis to a much larger and complete
set of physical parameters necessary to characterise the physical properties of the target galaxies.
Although a complete 
understanding of the far infrared emission of galaxies requires a comparison of observations with models, this empirical approach has the 
advantage of being free from any model dependent assumption and is
thus particularly useful for identifying the main parameters governing the dust emission of galaxies. A detailed comparison of the data with models 
will be presented in upcoming communications.\\

  
\section{The sample}

The $Herschel$ Reference Survey (HRS; Boselli et al. 2010a) provides us with an ideal sample for studying the relationship 
between the infrared colours and the physical properties of nearby galaxies spanning a large range in morphological type and luminosity.
The HRS is a K-band selected, 
volume limited (15 $<$ $D$ $<$ 25 Mpc), complete sample of galaxies at high galactic latitude ($b$ $>$ $+$55$^o$; $A_B$ $<$ 0.2, to avoid cirrus contamination) composed of
322 objects\footnote{With respect to the original sample given in Boselli et al. (2010a), we removed the galaxy HRS 228 whose new redshift indicates it as a background object.
We also revised the morphological type for three galaxies that moved from the early- to late-type class: NGC 5701, now classified as Sa, and NGC 4438 and NGC 4457 now Sb.}. 
Distances ($D$) have been determined assuming galaxies in Hubble flow with
$H_0$ = 70 km s$^{-1}$ Mpc$^{-1}$, outside the Virgo cluster. Within Virgo,
galaxies are taken at fixed distance according to the subgroup membership as indicated in Gavazzi et al. (1999).\\

The present analysis is focused on the far infrared (60 $\mu$m $\leq$ $\lambda$ $\leq$ 500 $\mu$m) properties of the 
late-type objects included in the sample. Several HRS galaxies belong to the Virgo cluster.
To avoid any possible second order effect related to the perturbations induced by the cluster 
environment, we restrict our analysis to those objects with a normal HI gas content (149 objects). 
There are indeed indications that the cold dust component of the interstellar medium of galaxies is removed during their interactions
with the hostile cluster environment (Boselli \& Gavazzi 2006; Cortese et al. 2010a, 2010b, 2012). We assume as normal, unperturbed objects those with an HI-deficiency parameter $HI-def$ $\leq$ 0.4,
where $HI-def$ is defined as the difference in logarithmic scale between the expected and observed HI mass of a galaxy of given angular size and morphological type 
(Haynes \& Giovanelli 1984). HI-deficiencies for all the target galaxies have been measured using the recent calibrations of Boselli \& Gavazzi (2009).

The statistical properties of galaxies belonging to the HRS are extensively described in Boselli et al. (2010a). To summarise, the objects analysed in this work
span a large range in morphological type (from Sa to Sm-Im-BCD), stellar mass (8.5 $\lesssim$ log $M_{star}$ $\lesssim$ 11 M$_{\odot}$), HI mass (7.5 $\lesssim$ log MHI $\lesssim$ 10.5
M$_{\odot}$), infrared (7.2 $\lesssim$ log $L_{60 \mu m}$ $\lesssim$ 10.2 L$_{\odot}$) and radio luminosity (20 $\lesssim$ log $L_{Radio}$ $\lesssim$ 22.5 W Hz$^{-1}$) and
star formation rate (0.1 $\lesssim$ $SFR$ $\lesssim$ 10 M$_{\odot}$ yr$^{-1}$). Their metallicity is
in the range 8.3 $\lesssim$ 12+log(O/H) $\lesssim$ 8.8, where the mean metallicity of the Milky Way 
in the solar neighborhood is 8.67 (Rudolph et al. 2006), and that of the LMC and SMC is 8.40 and 8.23, respectively (Korn et al. 2002; Rolleston et al. 2003).
The statistical properties of the dust component of the sample galaxies are described in a dedicated publication (Cortese et al. 2012).


\begin{table}
\caption{Completeness of the multifrequency data}
\label{Tab}
{
\[
\begin{tabular}{ccc}
\hline
\noalign{\smallskip}
Data			& N. obj. & Note\\
\hline
Complete~sample		& 149 	& 1 \\
SPIRE			& 146 	& 2 \\
IRAS			& 128 	& 3 \\
SFR			& 138	& 4 \\
b			& 138 	& 4 \\
$\Sigma(H\alpha$)	& 110   & 4 \\
$\mu_e(H)$		& 149	&   \\
12+log(O/H)		& 126	& 5 \\
A(H$\alpha$)		& 126	& 6 \\
A(FUV)			& 114	& 7 \\
\noalign{\smallskip}
\hline
\end{tabular}
\]
}
Notes: \\
1: late-type galaxies with an HI-deficiency $\leq$ 0.4.\\
2: detected in the three SPIRE bands.\\
3: detected by IRAS at 60 and 100 $\mu$m.\\
4: determined from H$\alpha$ data accurately corrected for [NII] contamination and dust extinction as described in Boselli et al. (2009). Galaxies hosting an AGN 
are not included.\\
5: with a metallicity measured using the same calibration (Hughes et al. 2012).\\
6: with a direct measure of the Balmer decrement from integrated spectra (Boselli et al. 2012).\\
7: with $A(FUV)$ directly measured from the far infrared to FUV flux ratio as described in Cortese et al. (2008).\\
\end{table}

\section{The data}

\subsection{$Herschel$/SPIRE data}

The HRS galaxies have been observed with the SPIRE instrument (Griffin et al. 2010) in the three bands at 250, 350 and 500 $\mu$m
as part of a SPIRE guaranteed time key project the $Herschel$ Reference Survey (Boselli et al. 2010a). Eighty three of the HRS galaxies lie in the footprint 
of the $Herschel$ Virgo Cluster Survey (HeViCS; Davies et al. 2010), an open time key project aimed at covering with PACS 
and SPIRE $\sim$ 60 sq. deg. of the Virgo cluster region. We took the SPIRE data of these 83 targets from HeViCS.

239 HRS objects plus 4 galaxies in HeViCS targeted by the HRS during the Herschel Science Demonstration Phase,
out of 322 galaxies of the HRS, have been observed using the SPIRE scan-map mode with a nominal scan speed of 30"/sec. Late-type galaxies analysed in this work have been
observed with three pairs of cross-linked scan maps to reach a pixel-by-pixel rms of $\sim$ 7, 8, 8 mJy/beam. To secure the detection of the dust associated with the extended HI disc,
all galaxies have been mapped at least up to 1.5 times their optical diameter. Galaxies with optical diameters smaller than $\simeq$ 180 arcsec were observed
using the small scan-map mode providing homogeneous coverage on a circular area of $\sim$ 5'. Larger galaxies have been observed in scan map mode, with
typical map sizes are  8'$\times$8', 12'$\times$12' and 16'$\times$16'.
The remaining 83 galaxies\footnote{Only 23/83 HeViCS galaxies match the conditions on the morphological type and the HI-deficiency parameter mentioned 
in the previous section. They are thus only 15 \% ~(23/149; see Table \ref{Tab}) of the objects analysed in this work.} 
located within the HeViCS footprint have been observed using the PACS/SPIRE parallel scan-map mode with a scan speed of 60"/sec
in four 16 sq.deg. different fields. 

Both the HRS and HeViCS data have been reduced using the map making pipelines developed within the Extragalactic Science Working Group (SAG2), extensively described 
in Bendo et al. (2011) and Davies et al. (2011). Data have been processed up to Level-1 using the standard SPIRE pipeline.
We use the BriGAdE method (Smith et al. in preparation) to remove the temperature drift and bring all
bolometers to the same level to secure the best baseline removal when temperature variations are present.
The different scans were then combined using the standard SPIRE making pipeline. The resulting images have pixel sizes of 6, 8 and 12 arcsec with FWHM of
18.2", 24.5" and 36.0" at 250, 350 and 500 $\mu$m, respectively.

Flux densities in the three SPIRE bands have been extracted using concentric elliptical apertures adapted 
to match the galaxy shape on the plane of the sky and to avoid unwanted contaminating sources (background objects, nearby companions, Galactic cirrus)
in the annulus selected for the sky determination, as extensively described in Ciesla et al. (2012). 
The maximal elliptical aperture is taken, whenever possible, at 1.4 times the optical radius 
of the galaxy (measured at the 25 B mag arcsec$^{-2}$), while the background in circular annuli at larger distances. For point like sources (2 objects matching the selection
criteria), flux densities are measured directly from the time line data, following the prescription proposed by G. Bendo and presented in the SPIRE 
manual\footnote{$http://$Herschel$.esac.esa.int/twiki/pub/Public/SpireCalibrationWeb/SPIREPhotometryCookbook_jul2011_2.pdf$} . Uncertainties are measured following 
the prescription of Boselli et al. (2003).
Typical uncertainties on the SPIRE flux densities of late-type galaxies due to the flux extraction procedure 
are of the order of 5.9, 7.6 and 10.5 \% ~ at 250, 350 and 500 $\mu$m, respectively (Ciesla et al.,
2012). Fifteen galaxies have independent data since they have been observed during both the HRS and the HeViCS survey. The median difference in their flux densities is 2, 2 and
3 \% at 250, 350 and 500 $\mu$m, respectively.
The uncertainty on the absolute calibration is 7 \%. This uncertainty should be added quadratically to the uncertainty on the extracted flux density of 
each single object.

The SPIRE pipeline transforms measured fluxes into monochromatic flux densities assuming that the dust emissivity of the emitting source changes as $\nu^{-1}$. 
It also assumes that the sources are point-like. Ciesla et al. (2012) transformed monochromatic flux densities from point-like to extended sources following the prescription given in
the SPIRE manual 
and described in Bendo et al. (2011). These monochromatic flux densities should be also corrected to take into account that the
spectral energy distribution of the observed galaxies in this spectral domain is better represented by a modified black body of emissivity $S(\nu,T) = B(\nu,T)\nu^{\beta}$
with $\beta$ $\sim$ 1.5-2 and $T$=15-30 K for $\beta$=2 and $T$=20-50 K for $\beta$=1.5 (see below).
For these ranges of $\beta$ and $T$ the corrections for extended sources such as those analysed in this work
are significantly smaller ($\leq$ 2 \%) than the uncertainty on the absolute calibration (7 \%) or on the measure of the flux density 
(6-11 \%), as indicated in Ciesla et al. (2012).
Even considering that a fraction of the observed flux of late-type galaxies comes from unresolved sources
within the galaxies (nucleus, HII regions...), for which larger corrections should be used, 
the mean colour corrections that should be applied to the total flux density of extended galaxies is $\lesssim$ 6.5 \% (excluding absolute calibration), thus smaller than the mean uncertainty
on the data. We thus decided not to apply any colour correction to the data.
Only detected sources are considered in the following analysis (146 objects).

\begin{table*}
\caption{Spearman correlation coefficients of the colour-colour relations (Fig. \ref{colours})}
\label{Tabcolcol}
{
\[
\begin{tabular}{cccccc}
\hline
\noalign{\smallskip}
Y-colour	& $S60/S100$	& $S60/S250$	& $S100/S250$	& $S250/S350$	& $S250/S500$	\\
\hline
$S60/S250$	& 0.86		&		&		&		&		\\
$S100/S250$	& 0.65		& 0.94		&		&		&		\\
$S250/S350$	& 0.30		& 0.55		& 0.65		& 		&		\\
$S250/S500$	& 0.28		& 0.52		& 0.63		& 0.98		&		\\
$S100/S500$	& 0.55		& 0.85		& 0.94		& 0.86		& 0.85		\\
\noalign{\smallskip}
\hline
\end{tabular}
\]
}
Notes: only high quality detections are considered (see Ciesla et al. 2012).
\end{table*}

\begin{table*}
\caption{Spearman correlation coefficients of the relations between the colour indices and the physical parameters (Fig. \ref{fisici})}
\label{Tabfisici}
{
\[
\begin{tabular}{cccccccc}
\hline
\noalign{\smallskip}
Y-colour	& log $SFR$		& log $b$	& log $\Sigma(H\alpha)$		& $\mu_e(H)$		& 12+log(O/H)	& $A(H\alpha)$	& $A(FUV)$	\\
Units		& M$_{\odot}$ yr$^{-1}$	&		& erg s$^{-1}$ kpc$^{-2}$		& AB mag arcsec$^{-2}$	&		& mag		& mag	\\
\hline
$S60/S100$	&-0.03			& 0.48		& 0.24					& -0.09			& -0.18		& -0.20		& 0.11	\\
$S60/S250$	& 0.01			& 0.44		& 0.44					& -0.24			& -0.05		& -0.14		& 0.29	\\
$S100/S250$	& 0.01			& 0.36		& 0.53					& -0.31			& 0.01		& -0.08		& 0.39	\\
$S100/S500$	& 0.14			& 0.21		& 0.65					& -0.46			& 0.29		& 0.08		& 0.55	\\
$S250/S350$	& 0.35			& -0.09		& 0.70					& -0.62			& 0.56		& 0.28		& 0.70	\\
$S250/S500$	& 0.29			& -0.10		& 0.70					& -0.61			& 0.55		& 0.26		& 0.68	\\
\noalign{\smallskip}
\hline
\end{tabular}
\]
}
Notes: only high quality detections are considered (see Ciesla et al. 2012).
\end{table*}

   \begin{figure}
   \centering
   \includegraphics[width=8cm]{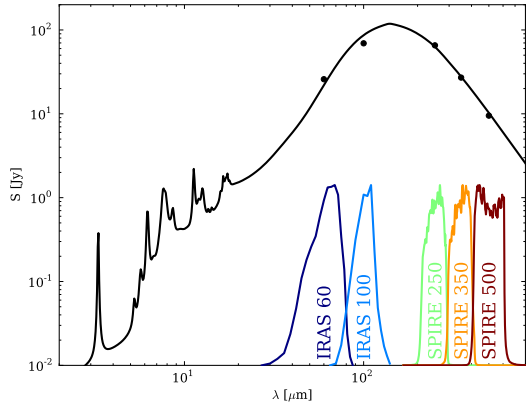}
   \caption{The infrared (2-800 $\mu$m) spectral energy distribution of the spiral galaxy M100 (NGC 4321). The solid line is, among the Draine \& Li (2007) dust models, 
   the one better fitting the observed data (filled dots). The IRAS 60 and 100 $\mu$m bands sample the dust emitting at wavelengths shorter than the 
   wavelength of the peak ($\sim$ 150 $\mu$m) and are thus sensitive to the emission of the warm dust. The SPIRE bands at 250, 350 and 500 $\mu$m
   cover the Rayleigh-Jeans tail of the dust emission (cold dust).
   }
   \label{SED}%
   \end{figure}

\subsection{Ancillary data and the derived physical parameters}

The SPIRE data have been combined with the IRAS 60 and 100 $\mu$m flux densities available for $\sim$ 90 \%~ of the late-type galaxies of the sample
taken from a large variety of sources (Sanders et al. 2003; Moshir et al. 1990; Thuan \& Sauvage 1992; Soifer et al. 1989; Young et al. 1996)
and collected on NED. These are integrated values thus directly comparable to the SPIRE flux densities
used in this work. The typical uncertainty of the IRAS flux densities is of the order of 15 \%. We do not apply any colour correction to the IRAS data.\\

Different sets of multifrequency data are required to characterise the physical properties of the interstellar medium and their relationships with
the dust emission properties in the far infrared spectral domain. 
The intensity of the general interstellar radiation field, i.e. the intensity of the radiation emitted by the bulk of the stellar component,
is quantified by means of the H-band effective surface brightness $\mu_e(H)$ (in mag arcsec$^{-2}$), defined as the mean surface brightness within the effective radius 
(radius including half of the total stellar light; see Gavazzi et al. (2000)). H-band magnitudes are available for all the target galaxies
from 2MASS (Jarrett et al. 2003) or from our own observations (GOLDMine; Gavazzi et al. 2003).

UV GALEX data, necessary to quantify the UV attenuation and the activity of star formation, have been taken
during the GALEX Virgo cluster survey (GUViCS, Boselli et al. 2011) and a dedicated Cycle 6 GALEX legacy program (Cortese et al. in preparation).
H$\alpha$+[NII] narrow band imaging data, recently obtained during a  
narrow band imaging survey (Boselli et al. in preparation), are used to quantify the intensity of the ionising radiation ($\Sigma(H\alpha)$) and the ongoing activity 
of star formation ($SFR$). FUV GALEX and H$\alpha$+[NII] narrow band imaging data are corrected for dust extinction (and [NII] contamination)
as described in Boselli et al. (2009).
Briefly, the H$\alpha$+[NII] narrow band imaging data are corrected for [NII] contamination and dust extinction using integrated spectroscopy data ($R$ $\sim$ 1000)
obtained for 126/146 of the sample galaxies at the OHP 1.93 m telescope (Gavazzi et al. 2004; Boselli et al. 2012). 
Standard recipes are used for those galaxies without spectroscopic data.

The Balmer decrement, here expressed as $A(H\alpha)$, is an independent, direct measure of the opacity of the interstellar medium.
$A(FUV)$, the attenuation in the FUV GALEX band at 1539 \AA, has been determined using the far infrared to FUV flux ratio following the 
prescription of Cortese et al. (2008). These recipes are indicated for normal, star forming galaxies such as those belonging to the HRS, where 
the general interstellar radiation field, and not only the radiation emitted by the youngest, most massive stars, contributes to the heating of the dust.\\

Extinction corrected H$\alpha$ fluxes and FUV flux densities are converted into star formation rates $SFR$ using the standard calibration of 
Kennicutt (1998). We assume an escape fraction of zero and a fraction of ionising photons absorbed by dust before ionising the gas of zero ($f$=1).
Although unphysical (see Boselli et al. 2009), this choice has been done to allow a direct comparison with the results obtained in the literature 
using other star formation rates determined using
H$\alpha$ data. These works generally generally assume $f$=1. Our most recent results indicate that $f$ $\sim$ 0.6 (Boselli et al. 2009).\\
H$\alpha$ fluxes are also used to quantify the intensity of the UV ionising radiation using the H$\alpha$ surface brightness $\Sigma(H\alpha$), where the 
H$\alpha$ emission is supposed to be as extended as the optical disc. This assumption is reasonable for unperturbed galaxies such as those analysed in this work (Boselli \&
Gavazzi 2006).

Near infrared imaging data combined with
UV GALEX and H$\alpha$+[NII] imaging data, are also used to quantify different direct tracers of the star formation history of the galaxies. 
This is done by measuring the birthrate parameter $b$ (Kennicutt et al. 1994), that in a closed box model can be defined as in
Boselli et al. (2001):

\begin{equation}
{b = \frac{SFR}{<SFR>}=\frac{SFR t_0 (1-R)}{M_{star}}}
\end{equation}

\noindent
with $t_0$ the age of the galaxy (13 Gyr) and $R$ the returned gas fraction, here assumed to be $R$=0.3 (Boselli et al. 2001). 
The total stellar mass $M_{star}$ is estimated using the H-band data and recently determined colour-dependent recipes (Boselli et al. 2009).
As defined, 

\begin{equation}
{b \propto \frac{L(H\alpha)}{L(H)}}
\end{equation} 

\noindent
measures the ratio of the ionising (photons with $\lambda$ $<$ 912 \AA) to non ionising ($\lambda$ = 1.65 $\mu$m) 
radiation and is thus a direct tracer of the hardness of the interstellar radiation field. Galaxies with a $b$ parameter $>$ 1 have 
a present day star formation activity more important than their mean star formation activity since their birth. They are 
characterised by very blue colours and thus have hard interstellar radiation fields.\\
The birthrate parameter is proportional to the specific star formation rate $SSFR$ defined as (Brinchmann et al. 2004):

\begin{equation}
{SSFR = \frac{SFR }{M_{star}} = \frac{b}{t_0 (1-R)}}
\end{equation}

\noindent
This parameter is also important since it is often used to discriminate the far infrared properties of galaxies in blind infrared cosmological surveys 
such as H-GOODS and H-ATLAS (e.g. Elbaz et al. 2011; Smith et al. 2011). We thus use either of the two parameters in the following analysis.\\

Integrated spectroscopy is also used to measure gas metallicities 12+log(O/H)
(Hughes et al., 2012) and Balmer decrement ($A(H\alpha)$; Boselli et al., 2012) 
for most of the late-type HRS galaxies. Depending on the availability of several
main emission lines, different calibrations have been adopted to
convert line emissions into 12+log(O/H). Following Kewley
\& Ellison (2008) we adopt the PP04 O3N2 calibration on the [NII] and [OIII] emission lines (Pettini \& Pagel 2004) as the base metallicity. We then 
determine the average oxygen abundance 12+log(O/H) for each galaxy. \\

HI gas data, available for almost the totality of the late-type galaxies of the sample, are used only to reject those objects that might have suffered any kind of
perturbation induced by the Virgo cluster environment. 
Table \ref{Tab} summarizes the completeness of the multifrequency data used for the present analysis.

   \begin{figure*}
   \centering
   \includegraphics[width=16cm]{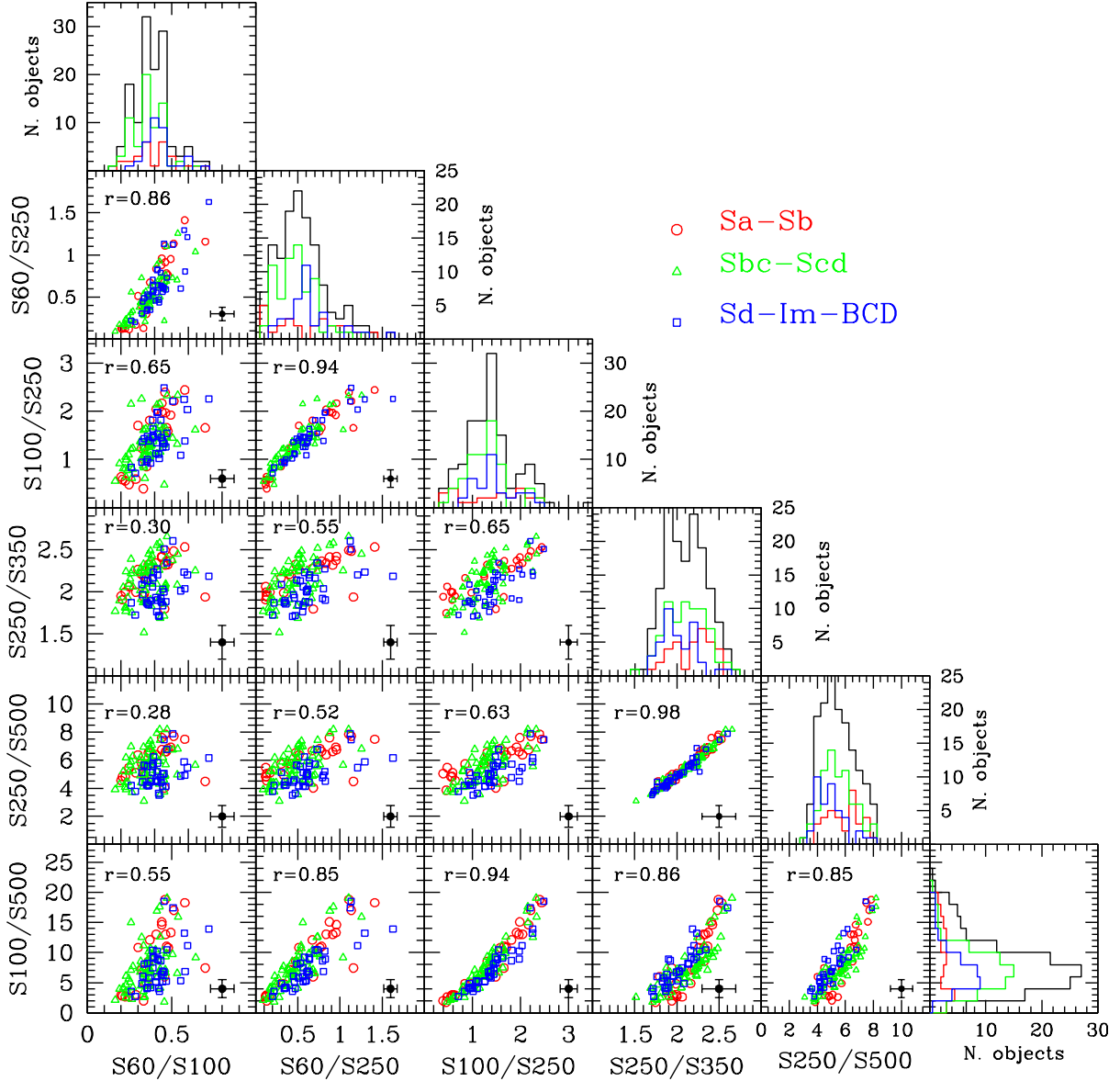}
   \caption{Far infrared colour-colour diagrams, where colours are defined as the ratio of the flux densities measured in two different bands. Galaxies are coded according to their
   morphological type, red open circles for Sa-Sb (34 objects), green empty triangles for Sbc-Scd (72) and blue open squares for Sd-Im-BCD (40). The typical error bar 
   is indicated with a black cross.
   }
   \label{colours}%
   \end{figure*}

   \begin{figure*}
   \centering
   \includegraphics[width=16cm]{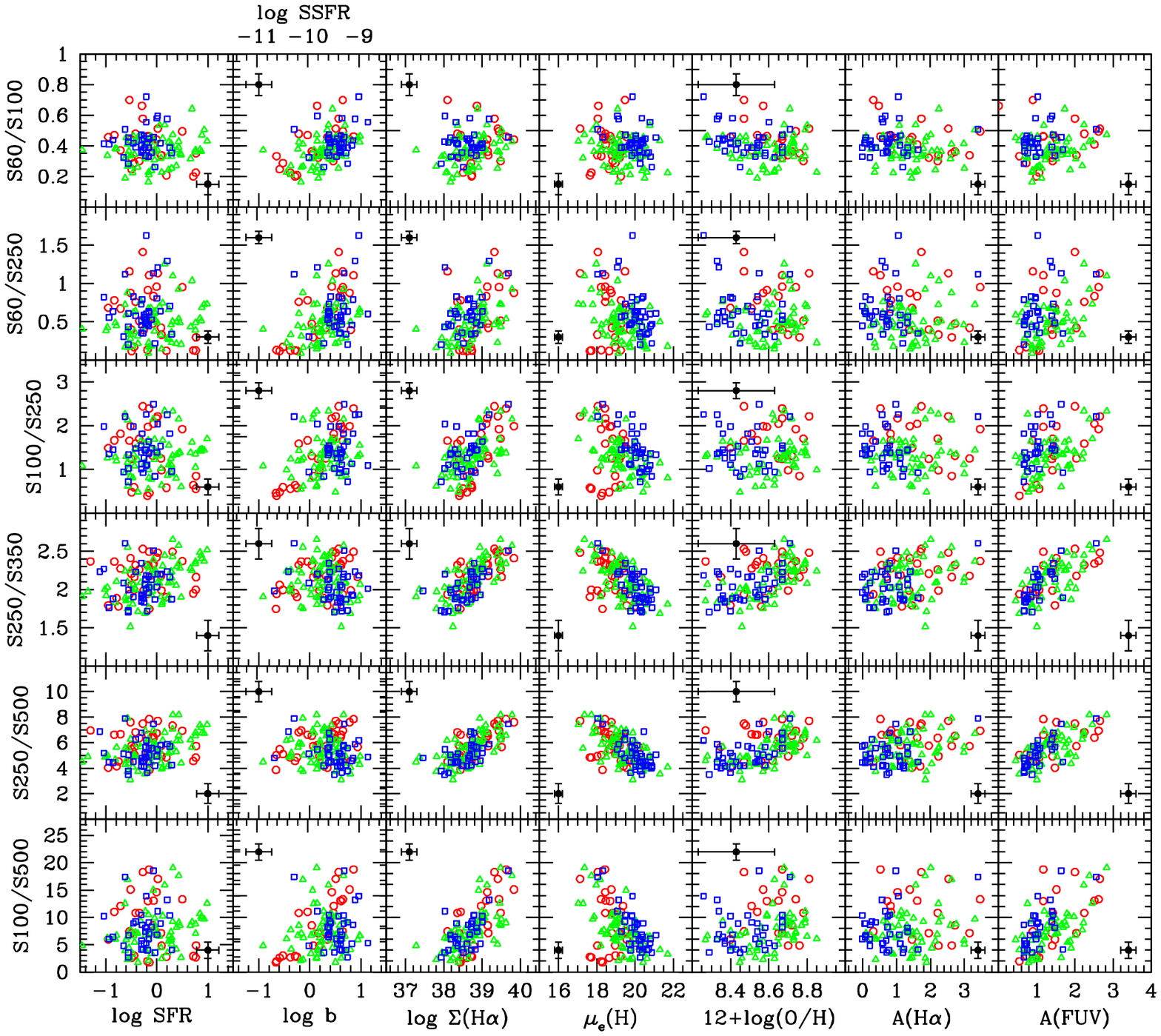}
   \caption{The relationship between the far infrared colour indices and different tracers of the physical properties of the interstellar medium, from left to right:
   first column: the logarithm of the star formation rate (in M$_{\odot}$yr$^{-1}$) measured as described in Boselli et al. (2009); second column: the logarithm of the 
   birthrate parameter $b$ (or the specific star formation rate $SSFR$); third column: the logarithm of the H$\alpha$ effective surface brightness (in erg s$^{-1}$ kpc$^{-2}$); fourth column: the H-band effective surface
   brightness (in AB mag arcsec$^{-2}$); fifth column: the metallicity index 12+log(O/H); sixth column: the Balmer extinction $A(H\alpha)$ (in magnitudes);
   seventh column: the FUV attenuation $A(FUV)$ (in magnitudes). 
   Red open circles for Sa-Sb, green empty triangles for Sbc-Scd and blue open squares for Sd-Im-BCD. The typical error bar is indicated with a black symbol.
   }
   \label{fisici}%
   \end{figure*}

\section{Far infrared colours}

\subsection{Far infrared colour indices}

Far infrared colours, here defined as the ratio of the flux densities measured in different far infrared bands, are useful quantitative tools to
characterise the properties of the interstellar dust in galaxies. Those analysed in this work 
include:\\
a) The widely used $S60/S100$ IRAS colour, often adopted as a direct tracer of the starburst activity of the
target galaxies. This index is sensitive to the emission of the warm dust component principally heated by young stars
(see Fig. \ref{SED}). 
Objects with $S60/S100$ $\geq$ 0.5 are generally considered as starbursts (Rowan-Robinson \& Crawford 1989). In the following 
we will refer to this colour index as to the {\it warm dust sensitive index}.\\
b) The colour indices $S60/S250$, $S100/S250$ and $S100/S500$ are sensitive to the relative weight of the warm and cold dust component 
since the peak of the emission of normal, star forming galaxies such as those analysed in this work lies in between 100 and 200 $\mu$m, thus in between the two
photometric bands used to define these indices (see Fig. \ref{SED}). They are related to the wavelength position of the peak of the dust emission.
We generally refer to these indices as the {\it dust peak sensitive indices}.\\
c) The SPIRE colour indices $S250/S350$, $S250/S500$ and $S350/S500$ are indices tracing the emitting properties of the coldest dust component
({\it cold dust sensitive indices}). 
Indeed they sample the Rayleigh-Jeans tail of the dust emission dominating in mass in normal galaxies (see Fig. \ref{SED}).

\subsection{Colour-colour diagrams}

Figure \ref{colours} shows the relationships between several far infrared colour indices for galaxies coded 
according to their morphological type. Table \ref{Tabcolcol} lists the Spearman correlation coefficients ($r$) of the relations.
Clearly, all colour indices are mutually related. The tightest correlations are present when the two plotted colour indices cover
the closest spectral bands. The correlations between the colour indices sensitive to the emission of the coldest dust component ($S250/S350$, $S250/S500$)
and the warm dust sensitive index $S60/S100$, however, are very marginal ($r$ $\sim$ 0.3).
We do not observe any strong systematic difference between galaxies of different morphological type.
The large dispersion in many colour-colour relations clearly indicates that the dust emission properties change significantly from galaxy to galaxy.
We thus try to investigate which physical parameter is at the origin of these variations.

   \begin{figure*}
   \centering
   \includegraphics[width=12cm]{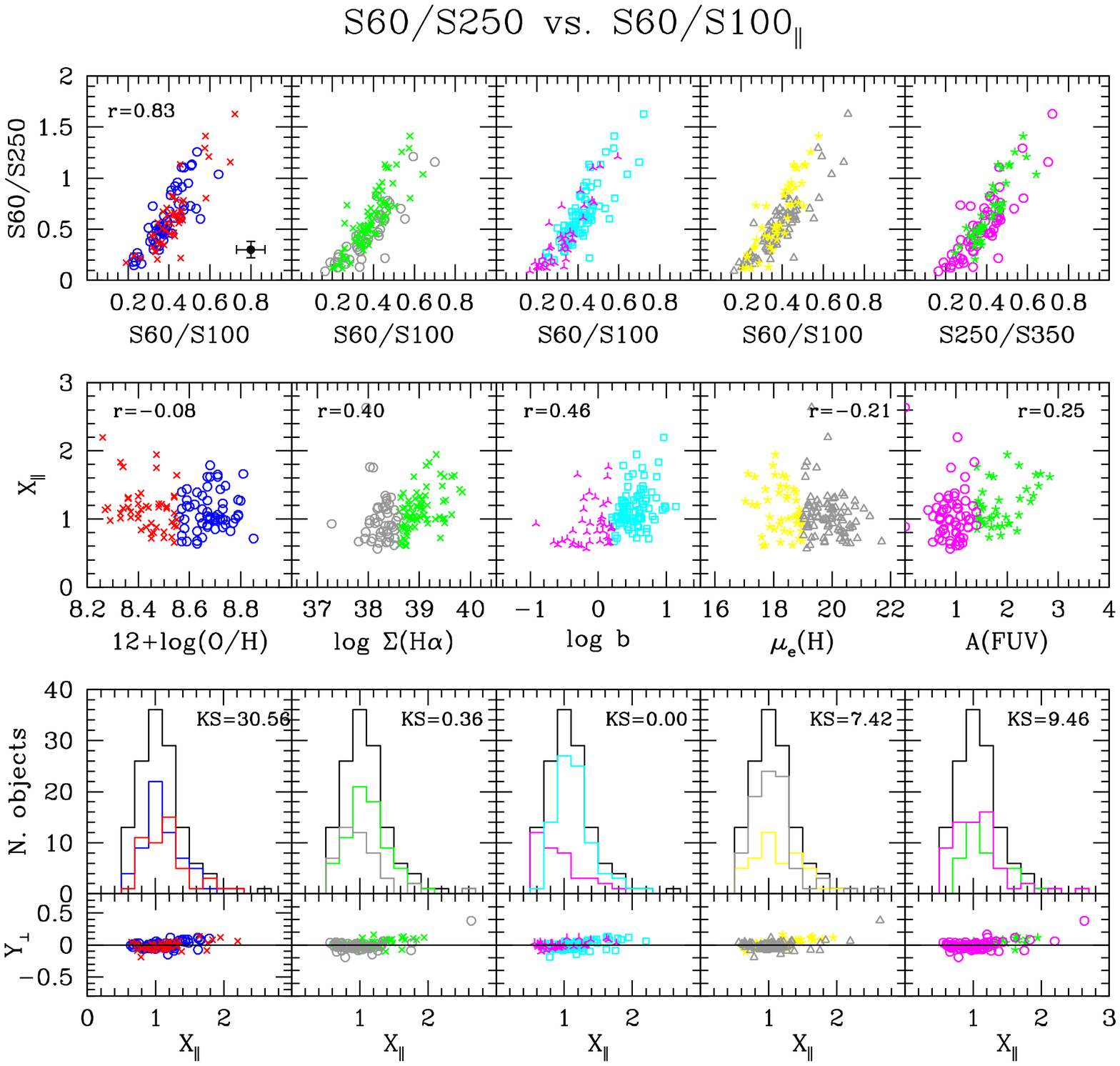}
   \caption{Upper panel: the $S60/S250$ vs. $S60/S100$ colour-colour relation with galaxies coded according to their physical parameters (from left to right) 
    as a function of the metallicity (red crosses for 12+log(O/H) $\leq$ 8.55, blue circles for 12+log(O/H) $>$ 8.55), H$\alpha$ surface brightness
   (grey circles for log $\Sigma(H\alpha$) $\leq$ 38.6 erg s$^{-1}$ kpc$^{-2}$, green crosses for log $\Sigma(H\alpha$) $>$ 38.6 erg s$^{-1}$ kpc$^{-2}$), 
   birthrate parameter (magenta three points stars for log $b$ $\leq$ 0.2, 
   cyan squares for log $b$ $>$ 0.2), H band effective surface brightness (yellow stars for $\mu_e(H)$ $\leq$ 19 AB mag arcsec$^{-2}$, grey triangles 
   for $\mu_e(H)$ $>$ 19 AB mag arcsec$^{-2}$) and UV attenuation (magenta circles for $A(FUV)$ $<$ 1.4 mag, green stars for $A(FUV)$ $\geq$ 1.4).
   Middle panel: the dependence of the position of galaxies along the $S60/S250$ vs. $S60/S100$ colour-colour relation ($X_{\parallel}$) on the different physical parameters. 
   Lower panel: the histogram of the distribution of galaxies along the colour-colour relation (upper)
   and the relation between the position of galaxies in the direction perpendicular ($Y_{\perp}$) and that along ($X_{\parallel}$) the colour-colour relation. 
   $r$ gives the Spearman correlation coefficient of the different relations, KS the probability that the two galaxy populations are driven by the same parent
   population (Kolmogorov-Smirnov test: for KS$\leq$ 5 the two galaxy populations are statistically different). The de-projection of the 
   $S60/S100$ vs. $S60/S250$ colour-colour relation is done using the measured linear best fit: $S60/S250$ = 2.64($\pm$0.16) $\times$ $S60/S100$ -
   0.45($\pm$0.06); $r$=0.83.}
   \label{histo60100X}%
   \end{figure*}

   \begin{figure*}
   \centering
   \includegraphics[width=12cm]{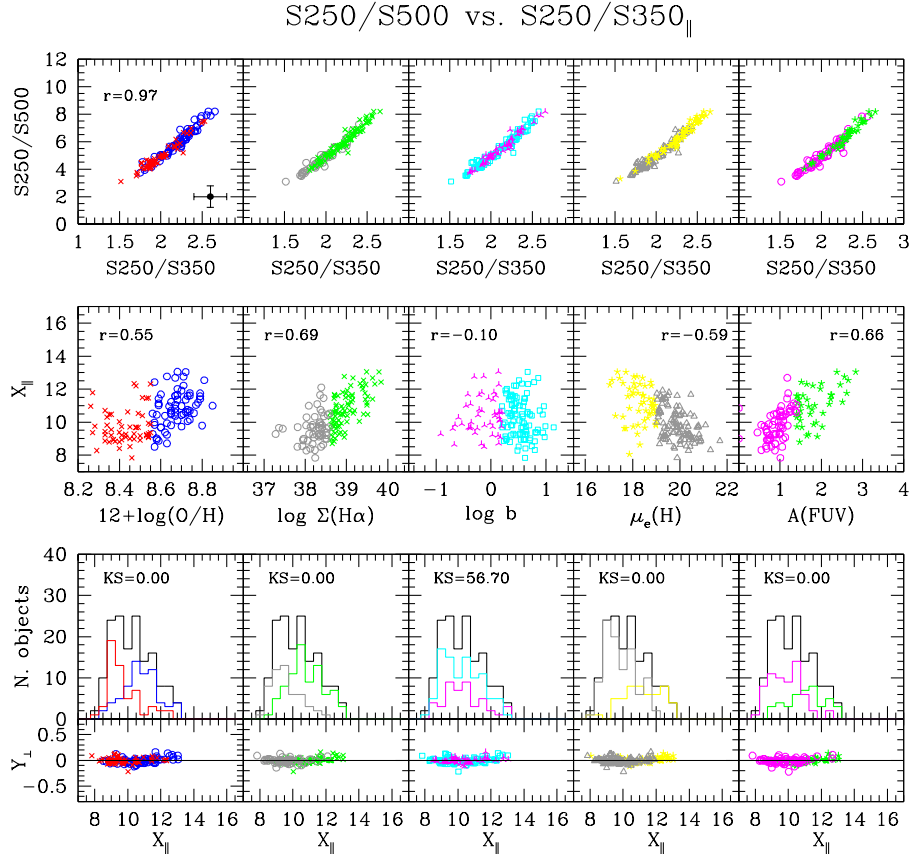}
   \caption{Upper panel: the $S250/S500$ vs. $S250/S350$ colour-colour relation with galaxies coded according to their physical parameters as in Fig. \ref{histo60100X}.
   Middle panel: the dependence of the position of galaxies along the $S250/S500$ vs. $S250/S350$ colour-colour relation ($X_{\parallel}$) on the different physical parameters.
   Lower panel: the histogram of the distribution of galaxies along the colour-colour relation (upper)
   and the relation between the position of galaxies in the direction perpendicular ($Y_{\perp}$) and that along ($X_{\parallel}$) the colour-colour relation. The de-projection of the 
   $S250/S500$ vs. $S250/S350$ colour-colour relation is done using the measured linear best fit: $S250/S500$ = 4.78($\pm$0.09) $\times$ $S250/S350$ - 4.59($\pm$0.19); $r$=0.97.}
   \label{histo250500X}%
   \end{figure*}

   \begin{figure*}
   \centering
   \includegraphics[width=12cm]{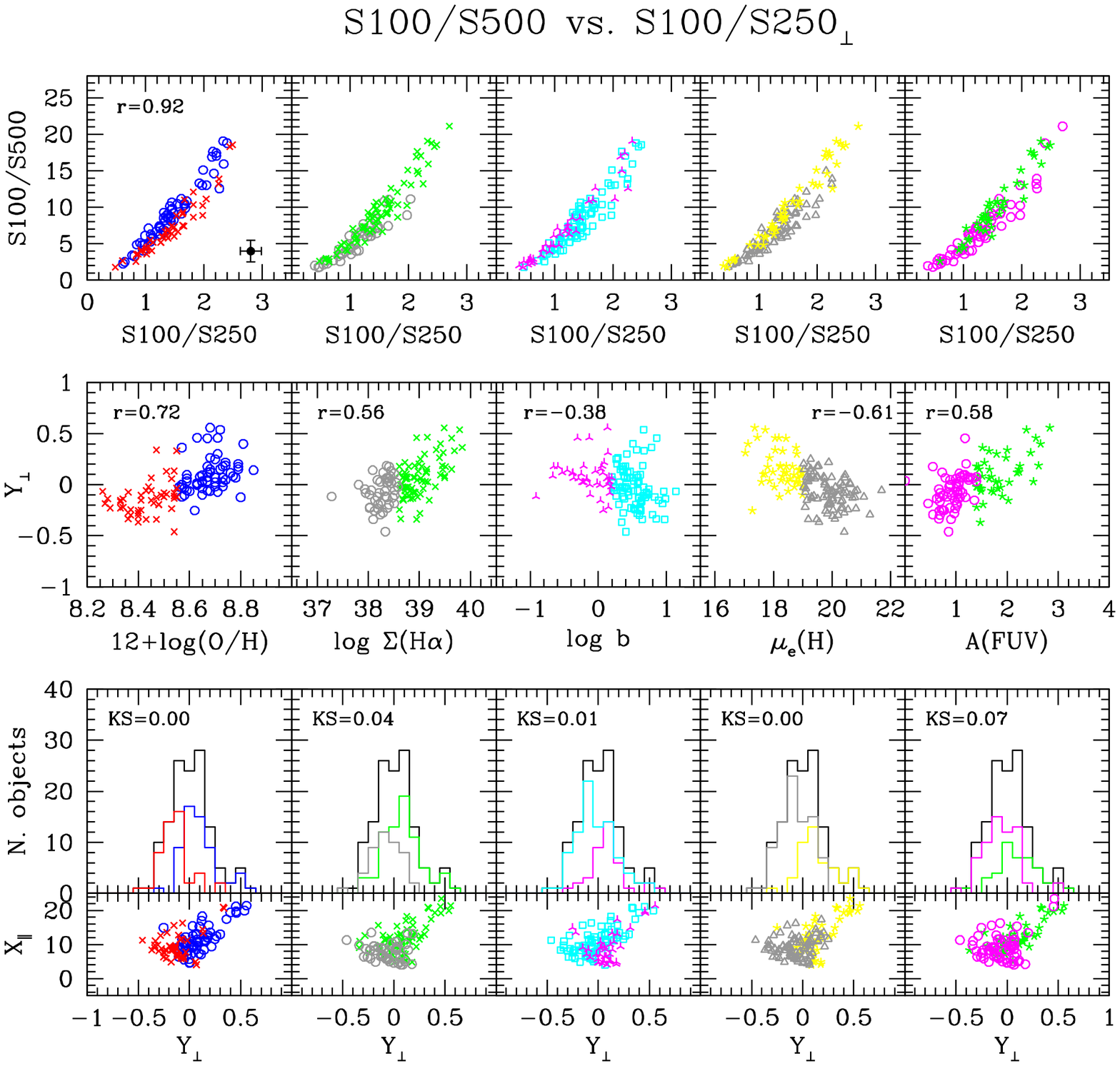}
   \caption{Upper panel: the $S100/S500$ vs. $S100/S250$ colour-colour relation for galaxies coded according to their physical parameters as in Fig. \ref{histo60100X}.
   Middle panel: the dependence of the position of galaxies in the direction perpendicular to the $S100/S500$ vs. $S100/S250$ colour-colour relation ($Y_{\perp}$)
   on the different physical parameters. 
   Lower panel: the histogram of the distribution of galaxies in the direction perpendicular to the colour-colour relation (upper)
   and the relation between the position of galaxies along ($X_{\parallel}$) and that in the direction perpendicular ($Y_{\perp}$) to the colour-colour relation. The de-projection of the 
   $S100/S500$ vs. $S100/S250$ colour-colour relation is done using the measured linear best fit: $S100/S500$ = 7.37($\pm$0.28) $\times$ $S100/S250$ - 2.23($\pm$0.41); $r$=0.92.}
   \label{histo100250Y}%
   \end{figure*}

   \begin{figure*}
   \centering
   \includegraphics[width=12cm]{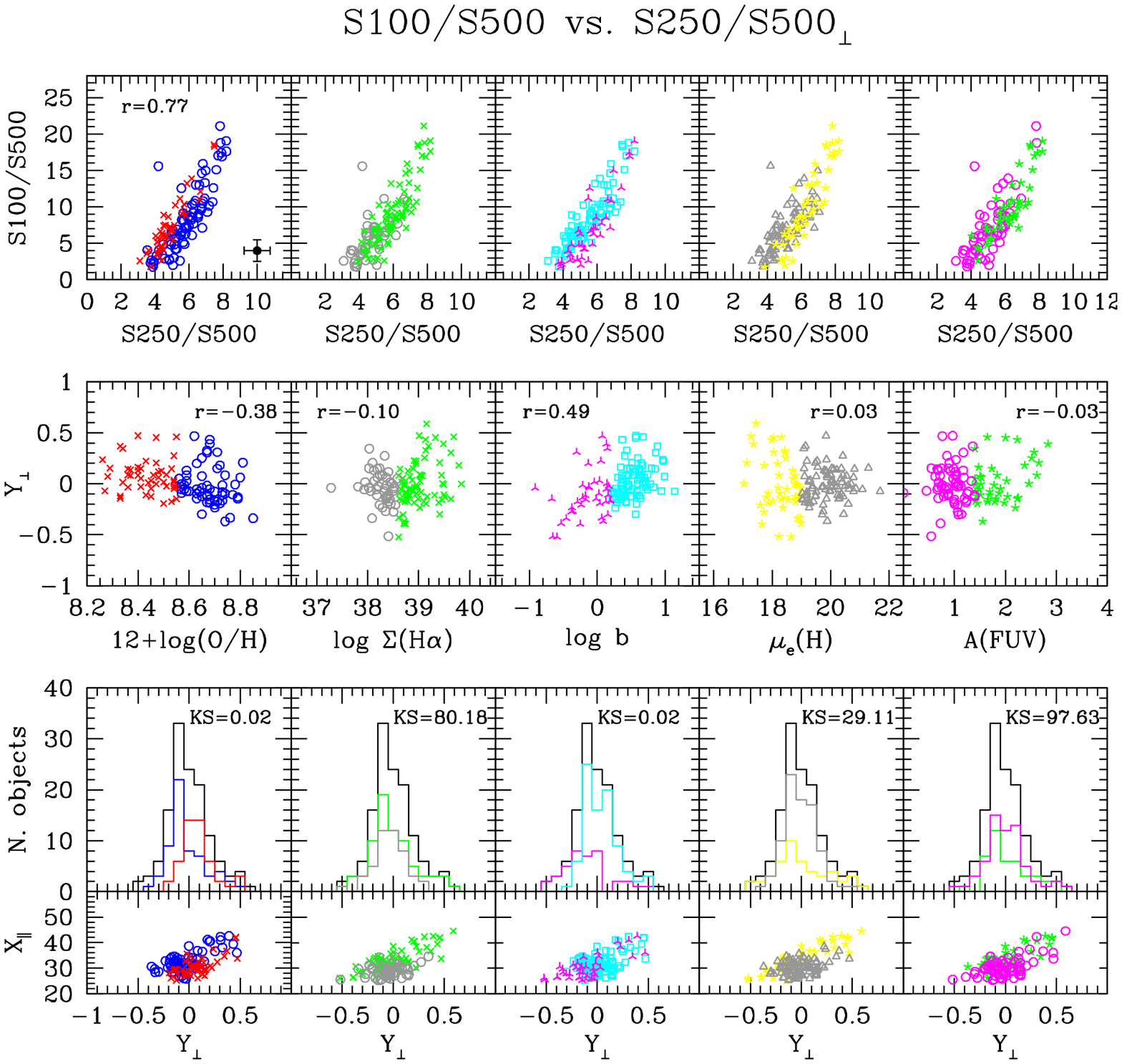}
   \caption{Upper panel: the $S100/S500$ vs. $S250/S500$ colour-colour relation for galaxies coded according to their physical parameters as in Fig. \ref{histo60100X}.
   Middle panel: the dependence of the position of galaxies in the direction perpendicular to the $S100/S500$ vs. $S250/S500$ colour-colour relation ($Y_{\perp}$) on the different
   physical parameters. 
   Lower panel: the histogram of the distribution of galaxies in the direction perpendicular to the colour-colour relation (upper)
   and the relation between the position of galaxies along ($X_{\parallel}$) and that in the direction perpendicular ($Y_{\perp}$) to the colour-colour relation. The de-projection of the 
   $S100/S500$ vs. $S250/S500$ colour-colour relation is done using the measured linear best fit: $S100/S500$ = 12.22($\pm$0.90) $\times$ $S250/S500$ - 23.43($\pm$2.32); $r$=0.77.}
   \label{histo250500Y}%
   \end{figure*}

\subsection{Colours vs. physical parameters}

The dust of the interstellar medium is heated by the interstellar radiation field. The energy absorbed by the dust grains is re-radiated in the infrared spectral domain.
It is thus expected that the dust emission properties are tightly related to the properties (intensity, hardness) of the interstellar radiation field. They can also depend on the
chemical composition of the emitting dust grains, which in turn can be related to the mean metallicity of the interstellar medium. To quantify these effects, we plot 
in Figure \ref{fisici} the relationships between the far infrared colour indices and different tracers of the physical properties of the interstellar medium. The Spearman correlation
coefficients of these relations are listed in Table \ref{Tabfisici}. These tracers are:\\
a) The star formation rate $SFR$, in solar masses per year.\\
b) The birthrate parameter $b$ that, as defined, is a tracer of the hardness of the incident interstellar radiation (hard for $b$ $\geq$ 1).\\
c) The H$\alpha$ surface brightness $\Sigma(H\alpha$), in erg s$^{-1}$ kpc$^{-2}$, which measures the intensity of the ionising radiation. Corrected for dust extinction,
this entity quantifies the surface density of the present day activity of star formation.\\
d) The H-band effective surface brightness $\mu_e(H)$, in AB mag arcsec$^{-2}$, which traces the intensity of the interstellar radiation field produced by the bulk of the stellar population. \\
e) The mean metallicity of the gaseous component, inferred from the 12+log(O/H) metallicity index. \\
f) The Balmer decrement $A(H\alpha)$, in magnitudes, which measures the attenuation of the ionising radiation within the gaseous component of the interstellar medium. \\
g) The attenuation of the non ionising UV radiation $A(FUV)$, in magnitudes, determined from the far infrared to UV flux ratio.\\

Figure \ref{fisici} shows that only a few far infrared colour indices analysed here are correlated with some of the physical entities used to characterise the properties of the
interstellar medium. The analysis of Fig. \ref{fisici} brings to the following conclusions:\\
1) Surprisingly, the warm dust sensitive index $S60/S100$ is not related to the direct tracers of the star formation activity ($SFR$, $\Sigma(H\alpha)$), while it is 
only (and marginally; $r$ = 0.48) correlated to the birthrate parameter $b$.
Galaxies still active in forming stars at the present epoch ($b$ $\ge$ 1) have, on average, $S60/S100$ flux density ratios ($S60/S100$ $\simeq$ 0.6) 
slightly larger than quiescent objects ($b$ $\le$ 1; $S60/S100$ $\simeq$ 0.4). This indicates that, whenever integrated values are used, the contribution of the 
warm dust component to the total emission of late-type galaxies is principally controlled by the history of star formation (or the hardness of the radiation field) 
rather than by the present day activity.\\
2) Poorly defined and dispersed relations ($r$ $\sim$ 0.3-0.6) are also observed between the peak sensitive $S100/S500$, $S100/S250$ and $S60/S250$ 
colour indices (in order of decreasing $r$) and the surface brightness of the ionising ($\Sigma(H\alpha$)) and non-ionising ($\mu_e(H)$) 
interstellar radiation field and the FUV attenuation ($A(FUV)$). Among these colour indices, only $S60/S250$ 
and $S100/S250$ ($r$ $\sim$ 0.4) correlate very mildly with the birthrate parameter. This evidence suggests that the position of the peak of the dust emission, or
in other words the mean temperature of the dust, is governed, as expected, by the general interstellar radiation field.\\
3) Strong correlations ($r$ $\gtrsim$ 0.6) are instead observed between the cold dust sensitive $S250/S350$ and $S250/S500$ colour indices 
and $\Sigma(H\alpha$), $\mu_e(H)$ and $A(FUV)$. Warmer colours are observed in those galaxies with higher radiation fields and higher extinction. 
The same colour indices are only barely related to the present day star formation activity ($SFR$; $r$ $\sim$ 0.3) and the metallicity ($r$ $\sim$ 0.55). Again, these plots 
indicate that the emission of the cold dust component is also controlled by the general interstellar radiation field and partly by the metallicity. 
What is surprising, however, is that the correlations with these cold dust sensitive indices are significantly stronger than with the peak sensitive indices
which are sampling a warmer dust component.\\
4) No evident correlations are observed between any of the far infrared colours and the attenuation $A(H\alpha)$. \\
5) We also do not observe any evident systematic differences among galaxies of different morphological 
type, maybe with the exception that intermediate type spirals (Sbc-Scd, green empty
triangles) covers a larger range in the parameter space than the other classes.\\

Figure \ref{fisici} shows that the far infrared colour indices can be correlated at the same time with apparently different tracers of the properties of galaxies. 
The most striking example are the strong correlations observed between the cold dust sensitive colour indices and $\Sigma(H\alpha$) and $\mu_e(H)$, the former tracing 
the surface density of the radiation emitted by newly formed stars, the latter that of the very evolved stars. Although measured using independent data, the different
physical parameters used in this analysis are sometime related with each other, as discussed in appendix A. The mutual relationships between the different physical parameters 
must be considered for a complete and coherent interpretation of Fig. \ref{fisici}.

To better understand the role of the different physical parameters in the definition of the various colour-colour relations plotted in Fig. \ref{colours}
we analyse in detail a few colour-colour diagrams selected to represent at the same time those colour indices sensitive to the warm dust component, to the position of the peak
of the dust emission, and the cold dust emission. These relations are shown in Fig. \ref{histo60100X}, \ref{histo250500X}, \ref{histo100250Y} and \ref{histo250500Y} (upper panel)
\footnote{For such a purpose here we analyse in detail only the less dispersed colour-colour relations shown in Fig. \ref{colours} because: i) only here
the linear fit (and its residual) can be accurately determined; ii) variations in the more dispersed colour-colour relations can be directly determined from
the inspection of Fig. \ref{colours} once galaxies are coded according to different physical parameters, as done in the Appendix B 
(see Fig. \ref{metal}, \ref{SHa}, \ref{bpar}, \ref{mueh}  and \ref{AFUV}). }.
To this aim, galaxies are divided into two subsamples and coded according to different physical parameters, namely the metal content 
(with a threshold for selecting galaxies taken at 12+log(O/H) = 8.55), the H$\alpha$ surface brightness (log $\Sigma(H\alpha$) = 38.6 erg s$^{-1}$ kpc$^{-2}$),
the birthrate parameter (log $b$ = 0.2), the H-band effective surface brightness ($\mu_e(H)$ = 19 AB mag arcsec$^{-2}$)
and the FUV attenuation ($A(FUV)$ = 1.4 mag). We choose these physical parameters because
they seem to be the most important in defining the observed relations in Fig. \ref{fisici}. The adopted thresholds have been chosen to split the sample in two sub-samples
each with approximately the same number of objects. For simplicity we focus here only on some representative colour-colour relations, referring the interested reader 
to appendix B for the rest of the colour-colour diagrams shown in Fig. \ref{colours}. 
To quantify the role of the different physical parameters in defining the observed 
colour-colour relations, we first calculate the the best fit of the relations shown in Fig. \ref{histo60100X}, \ref{histo250500X}, \ref{histo100250Y} and \ref{histo250500Y} (upper panel).
We then use the best fit to de-project these colour-colour relations, and then study how the obtained variables running 
along ($X_{\parallel}$) or perpendicular ($Y_{\perp}$) to these relations depend on the different physical parameters (middle panel). We then quantify the difference 
in the various subsamples of objects through a Kolmogorov-Smirnov test (lower panel). The position of galaxies along or in the direction perpendicular to that of these colour-colour 
diagrams indicate variations in their spectral energy distribution due to either a variation of the mean dust temperature or of the dust grain emissivity in the sampled photometric bands.

Figures \ref{histo60100X}, \ref{histo250500X}, \ref{histo100250Y} and \ref{histo250500Y} clearly show that galaxies coded with different symbols populate these colour-colour diagrams
following a well defined order. Interestingly, galaxies can be at the same time well mixed and clearly segregated in a given colour-colour diagram
according to their different physical parameters (indicated by different symbols and colour codes in the figures). An evident example is the 
mixed vs. segregated distribution of galaxies in Fig. \ref{histo250500X} when coded according to their birthrate parameter $b$ or H$\alpha$ surface brightness:
while galaxies with different $b$ populate the whole dynamic range of the $S250/S500$ vs. $S250/S350$ relation, those of high $\Sigma(H\alpha)$ have systematically warmer
infrared colours than objects with low ionising radiation fields.
All together, this is a further evidence that the emission properties of the dust are tightly related to the properties of the interstellar radiation field (hardness, intensity),
to the internal attenuation and to the mean metallicity of the interstellar medium in a quite complex manner which can change as a function of wavelength.\\

A detailed, comparative analysis of Figures \ref{fisici}, \ref{histo60100X}, \ref{histo250500X}, \ref{histo100250Y} and \ref{histo250500Y} 
at the same time confirms our previous findings and brings to light several new and interesting results:\\
1) The commonly used colour index $S60/S100$, tracer of starburst activity sensitive to the emission of the warm dust component, 
depends only very marginally on the star formation history of the galaxy $b$ (or equivalently, on the hardness of the interstellar radiation field)
and on the surface brightness of the youngest stars ($\Sigma(H\alpha)$; Fig. \ref{histo60100X}). Galaxies are indeed located along the $S60/S250$ vs. $S60/S100$ 
colour-colour relation according to the birthrate parameter $b$ ($r$=0.46) and $\Sigma(H\alpha$) ($r$=0.40).\\
2) Galaxies are distributed in the direction perpendicular to the colour-colour relations defined with indices sensitive to the position of the peak 
of the far infrared emission (Fig. \ref{histo100250Y}) according to all the physical parameters used in this work but $b$, where any correlation, if present, is very weak ($r$=-0.38). \\
3) The properties of the coldest dust component, as traced by the $S250/S350$ vs. $S250/S500$ colour index, do not depend on the birthrate parameter $b$ (magenta three points stars and 
cyan open squares are well mixed in Fig. \ref{histo250500X}). The distribution of galaxies along this colour-colour relation changes with $\Sigma(H\alpha$)
($r$=0.69), $\mu_e(H)$ ($r$=-0.59) and $A(FUV)$ ($r$=0.66) and only very marginally on metallicity ($r$=0.30).\\
4) In colour-colour relations that are sensitive to the cold dust and their wavelength peak, the perpendicular distance to these trends
are only moderately related to a galaxy's star formation history ($r$=0.49) and metallicity ($r$=-0.38).


\subsection{Comparison with a single modified black body emission}

   \begin{figure*}
   \centering
   \includegraphics[width=16cm]{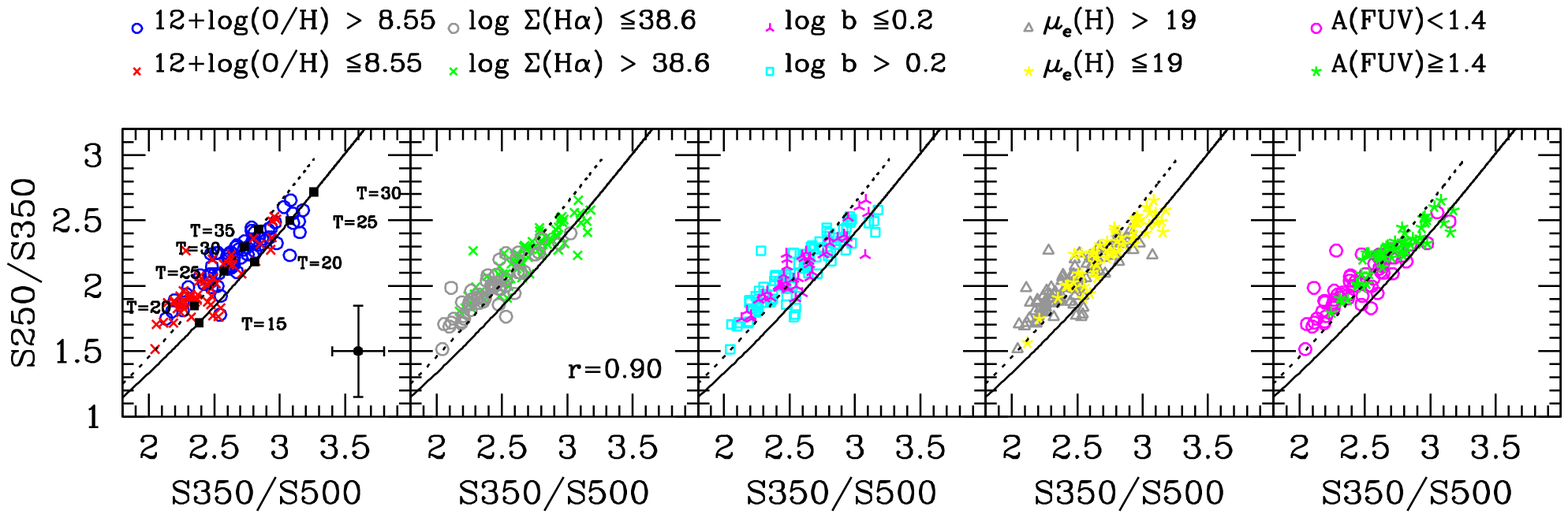}
   \caption{The relationship between the SPIRE colour-colour indices $S250/S350$ vs. $S350/S500$ for galaxies coded according to their different physical parameters as in Fig.
   \ref{histo60100X}. The colour-colour relation is compared to the expected relations 
   obtained for a modified black body with a grain emissivity index $\beta$ = 2 (solid line) and $\beta$ = 1.5 (dotted line). Black squares indicate different temperatures for the two modified
   black bodies (left panel). The Spearman correlation coefficient of this relation is $r$=0.90.  }
   \label{BBspire}%
   \end{figure*}

   \begin{figure*}
   \centering
   \includegraphics[width=16cm]{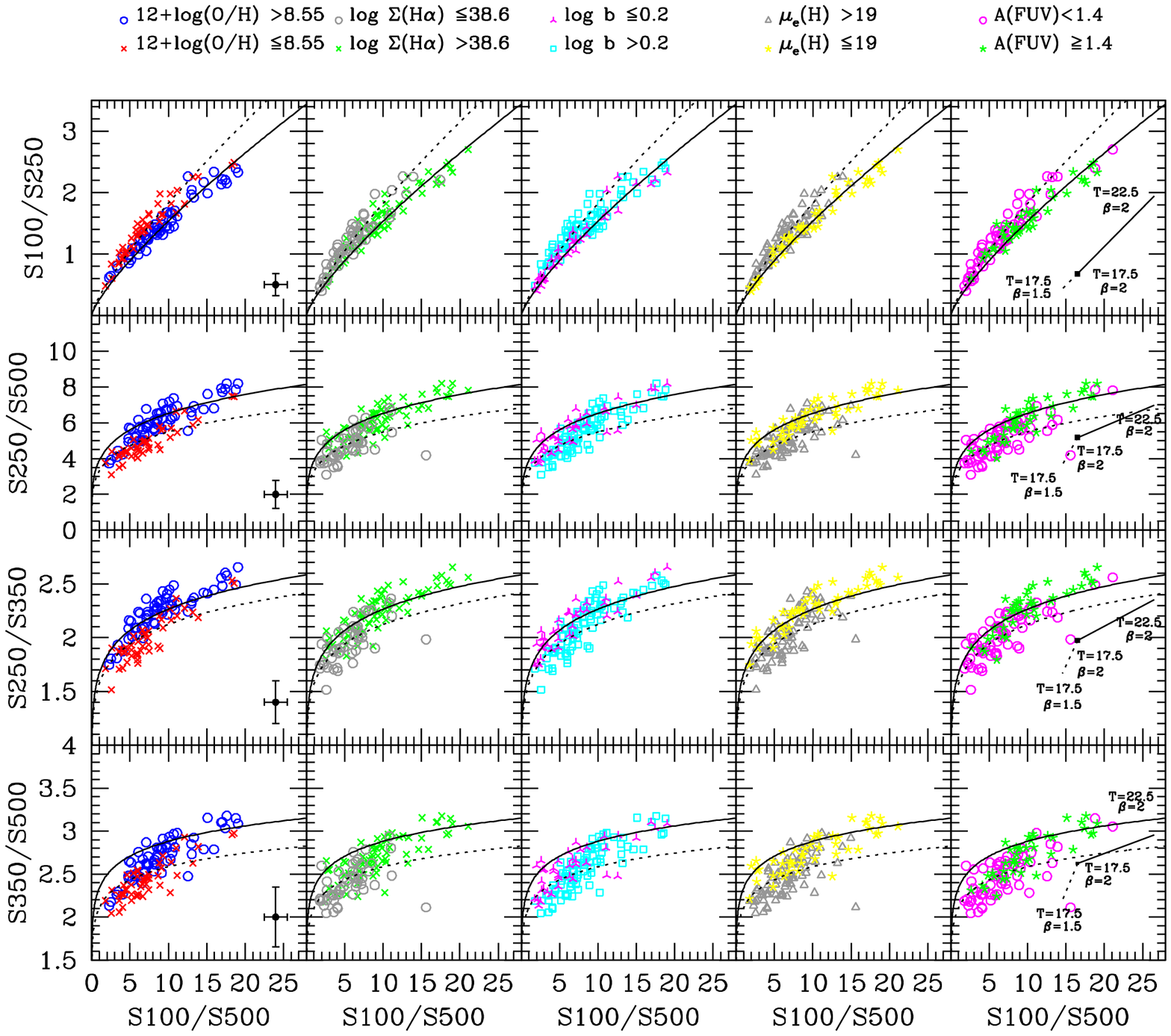}
   \caption{The comparison of relationships between different far infrared colour indices for galaxies coded according to their different physical parameters as in Fig.
   \ref{histo60100X}. The solid and dotted lines indicate the expected
   ratios for a single modified black body with grain emissivity $\beta$=2 and 1.5, respectively. The short solid and dotted lines in the last column show the
   expected variations when the temperature $T$ changes from 17.5 to 22.5 K for $\beta$=2 (solid line), and from 17.5 K and  $\beta$=2 to $T$=17.5 K
   and $\beta$=1.5 (dotted line). The black square is the expected value (shifted on the X-axis for clarity) for a black body of $T$=17.5 K and  $\beta$=2 .
   }
   \label{BB}%
   \end{figure*}

The relationships observed between the different far infrared colour indices can be due either to variations of the mean temperature of the emitting dust or to variations in the dust grain 
emissivity properties that might change in different types of galaxies. Assuming that dust grains are in thermal equilibrium with the radiation, which is probably the case for $\lambda$
$\geq$ 250 $\mu$m, 
the far infrared emission of galaxies is generally approximated by one (or more) modified black body $B(\nu,T)$, with a resulting emitted flux density: 

\begin{equation}
{S_{dust}(\nu,T) \propto \nu^{\beta}B(\nu,T)}
\end{equation}
 
\noindent
where $\beta$ is the grain emissivity index whose value ranges between $\sim$ 1.5 and $\sim$ 2 (Hildebrand 1983). In particular, a single modified black body with a temperature of the order of $\sim$
20 K is generally used to reproduce the rapid decrease of the far infrared emission observed at long wavelengths (Reach et al. 1995; Bianchi et al. 1999; Dunne et al. 2000; 
Dunne \& Eales 2001; Bendo et al. 2003; Draine 2003; Davies et al. 2011; Magrini et al. 2011). This assumption is not physical since it is not indicated to represent a continuum 
in the dust grain distribution in matter of size and temperature (Mathis et al. 1977, Draine \& Li 2007). However this simple analytical prescription is often used both in extragalactic and cosmological studies 
since it reproduces fairly well the observed far infrared spectral energy distributions of galaxies. Our new homogeneous and complete set of data, in particular those obtained by SPIRE in 
the spectral domain 250-500 $\mu$m, can be used to see whether this simple modified black body assumption is realistic. At the same time this dataset can help us 
to understand which of the two main parameters
regulating the dust emissivity, the grain emissivity index $\beta$ or the dust temperature $T$, is the main driver of the observed colour-colour far infrared relations and for their scatter.
To this aim we plot in Fig. \ref{BBspire} the SPIRE colour-colour indices $S250/S350$ vs. $S350/S500$ for galaxies coded according to the different physical parameters already used in the previous
figures and compare them to the expected values obtained for a modified black body with a grain emissivity index $\beta$ = 2 (solid line) and $\beta$ = 1.5 (dotted line). Clearly, the observed
relations can not be represented by a modified black body with a dust grain emissivity index of $\beta$ = 2, but are better reproduced when $\beta$ = 1.5. 
Systematic errors of the order of $\sim$ 15 \% on a colour are necessary to shift the data onto a $\beta$=2 single black body. This result seems thus robust vs. calibration
uncertainties, that are at present estimated to be of the order of $\simeq$ 7 \%~ for each band, or colour corrections ($\lesssim$ 6.5 \%). As expected, these relationships might just be due to a variation of the mean temperature of the dust grains
which, for a $\beta$ = 1.5, might range from $\sim$ 15 to $\sim$ 50 K (from $\sim$ 15 to $\sim$ 30 K for $\beta$=2). These values of $\beta$ and $T$
are consistent with that found in the Milky Way using COBE data by Reach et al. (1995)
or in other nearby galaxies observed in the submillimetre domain (Bianchi et al. 1998; 1999; Amblard et al. 2010; Hwang et al. 2010; Dunne et al. 2011; Planck Collaboration
 2011b). \\

The observed relationship between the two colour indices, however, has a slope slightly flatter than the one predicted by a single 
modified black body of fixed grain emissivity index. We can thus not exclude values of $\beta$ $>$ 1.5 in galaxies with the highest 
flux density ratios. Similarly, in the objects with the lowest flux density ratios, $\beta$ might be $<$ 1.5.
This evidence might indicate that both $\beta$ and $T$ vary along the 
sequence. 
Indeed, there is a quite strong degeneracy between $\beta$ and $T$
given that at these low temperatures a variation of $T$ of $\sim$ 5 K is almost equivalent to a variation of $\beta$ of $\sim$ 0.5.  
Furthermore, we have to remember that $\beta$ and $T$ might also be inversely correlated 
(Desert et al. 2008; Shetty et al. 2009; Veneziani et al. 2010; Anderson et al. 2010; Planck Collaboration 2011a; Bracco et al. 2011).\\

By identifying galaxies according to their physical properties we have shown that high flux density ratios are mainly observed in objects with strong ionising
radiation fields, which are generally also metal rich galaxies with strong non-ionising stellar radiation fields. If this single modified black body configuration is
valid, the grain emissivity index of metal poor objects with low ionising radiation fields is $\beta$ $\lesssim$ 1.5. 
It is worth mentioning, however, that all the previous considerations on the cold dust emission are valid if the contribution of the warm dust component 
to the far infrared emission ($\lambda$ $\geq$ 250 $\mu$m) is negligible, which is probably the case given that our sample is composed of normal, late-type galaxies and do not include strong
starbursts such as M82 or Arp 220 (Boselli et al. 2010a).

\noindent
The same analysis can be done by comparing other far infrared colour-colour relations defined for galaxies coded according to the same physical parameters (Fig. \ref{BB}).
Figure \ref{BB} confirms the trend observed in the previous figures with the different physical parameters.
Only the $S100/S250$ vs. $S100/S500$ colour-colour relation for metal poor galaxies or for objects with a high H$\alpha$ or H band surface brightness (upper row
in Fig. \ref{BB}) can be easily reproduced by a single black body with a grain emissivity parameter $\beta$=2.
It is however clear from both Fig. \ref{BBspire} and \ref{BB} that all the other colour-colour relations made using at least one cold dust sensitive colour index cannot be well reproduced by
a single modified black body. This evidence confirms that either the grain emissivity parameter $\beta$ or the temperature significantly change from galaxy to galaxy, as indeed expected given the
variety of physical conditions that characterise these objects. Fitting the infrared spectral energy distribution with different dust models might help us to make a further step in  
understanding the very nature of the emitting dust. This will be the topic of a future communication.

\section{Discussion}

The analysis of the far infrared colours of nearby late-type galaxies have been presented in the literature using data taken from the 
IRAS (e.g. Helou 1986) and Spitzer (Dale et al. 2005; 2007). 
This is the first dedicated work on the analysis of the far infrared colours of nearby late-type galaxies in the SPIRE/$Herschel$ bands.
Our results can thus be compared only to those obtained for a few nearby galaxies recently observed by $Herschel$.
Variations of the infrared colour indices with the H$\alpha$ and the H-band surface brightness similar to those presented in this work have been also observed on galactic scales over 
the disc of M81, M83, NGC 2403 (Bendo et al. 2011) and of M33 (Boquien et al. 2011).
The analysis of the Spitzer and $Herschel$ data of these well known galaxies confirmed that while the far infrared emission in the 70-160 $\mu$m spectral range is principally 
governed by the activity of star formation, the cold dust responsible for the emission at longer wavelengths is predominantly heated by the general interstellar radiation field. 
Variations of the far infrared colour indices as a function of metallicity have been also reported on galactic scales in M99 and M100 by Pohlen et al. (2010) and in unresolved objects by Boselli et al. (2010b).\\

The new observational evidence presented in this work puts a strong constraint for the study of the physical properties of the dust grains emitting in the far infrared spectral domain.
It shows that the properties of the dust emitting in this spectral domain are not universal but rather vary 
in galaxies due to their large range of parameter space. Qualitatively, the results discussed in the previous section can be interpreted as follow:\\
1) The colour-colour diagrams and the relationships between colours and the different physical parameters are tighter when SPIRE bands are used instead of the 60 and 100 $\mu$m IRAS bands.
This evidence might be related to a lower photometric quality of the IRAS data with respect to the new SPIRE observations, or to the larger wavelength range sampled by SPIRE (250-500 $\mu$m)
with respect to IRAS (60-100 $\mu$m).  \\
2) The weak dependence of $S60/S100$ on the birthrate parameter $b$ (Fig. \ref{fisici} and \ref{histo60100X}) is a further indication of a well known result, i.e. that
this colour index is related to the star formation history of galaxies. This tendency, however, is much less clear than that observed in other samples (e.g. Chapman et al. 2003) 
just because the dynamic range in the star formation history of mass-selected HRS galaxies is quite small (0.1 $\lesssim$ $SFR$ $\lesssim$ 10 M$_{\odot}$ yr$^{-1}$).
Indeed the increase of the $S60/S100$ ratio as a function of the infrared luminosity reported in Chapman et al. (2003) is evident only for total infrared luminosities 
$L_{TIR} \geq 10^{10}$ L$_\odot$, which roughly corresponds to the upper limit in the infrared luminosity sampled by the HRS galaxies (Boselli et al. 2010a).\\
3) The dependence of the SPIRE colour indices $S250/S350$, $S350/S500$ and $S250/S500$ on metallicity (Fig. \ref{fisici}) is probably a different indication
of the widely known submillimetre excess in metal poor, low surface brightness dwarf galaxies (e.g. Galametz et al. 2011). Indeed, these far infrared flux density ratios are 
low in metal poor objects characterised by a low FUV attenuation and a low ionising and non-ionising radiation field.
The observed variations in the colour-colour diagrams might be due to variations in the grain emissivity parameter $\beta$.
A low value of $\beta$ has been already proposed to explain the excess of the submillimetre 
emission at $\lambda$ $\geq$ 500 $\mu$m observed in metal poor, dwarf galaxies (Galliano et al. 2005, Bendo et al. 2006, Galametz et al. 2010, O'Halloran et al. 2010, 
Boselli et al. 2010b). Silicates and amorphous carbons dominate 
the emission at long wavelengths, so if their fraction changes, different values of $\beta$ might be expected (Compiegne et al. 2011).
Stellar evolution models, however, do not predict for a given metallicity strong differences in the relative production of C and Si for galaxies undergoing a different star formation
history that might justify a systematic difference in the abundance of silicate and carbonaceous dust grains.
Despite a large uncertainty in the dust production and destruction processes, models of dust formation in galaxies determined for different star formation histories 
do not predict a significant difference in the fraction of carbonaceous and silicate dust grains
(Dwek 1998; Calura et al. 2008). The excess of the cold dust emission in low luminosity, metal poor galaxies might have also other origins, as extensively discussed in
Galametz et al. (2011). These include a dependence of the dust grain emissivity with temperature (Dupac et al. 2003; Meny et al. 2007; Desert et al. 2008), variation 
of the grain emissivity with the fractal aggregation of individual amorphous grains inside molecular clouds (Paradis et al. 2009), grain coagulation (Bazell \& Dwek 1990),
overabundance of very small grains with grain emissivity $\beta$=1 in extreme environments (Lisenfeld et al. 2001; Zhu et al. 2009) or spinning dust (Ferrara \& Dettmar 1994).\\
4) The presence of parallel colour-colour relations for galaxies of different physical parameters in the diagrams done with 
flux density ratios sensitive to the position of the peak (defined by the different symbols in the relations shown in Fig. \ref{histo100250Y}) indicate that the peak of the dust emission in 
the spectral energy distribution, and thus the mean temperature of the dust, shifts from short to long wavelengths as a function of these physical parameters. \\
5) Far infrared colours ($\lambda$ $\geq$ 250 $\mu$m) change with the surface brightness of the ionising and non-ionising radiation, and the FUV dust attenuation  (Fig. \ref{fisici}, \ref{histo250500X}).
This result is expected since the integrated emission of galaxies 
includes both HII regions along the spiral arms, where the heating sources are mainly young and massive OB stars, and the more quiescent interarm regions where dust is principally heated by the
general interstellar radiation field. Although $A(FUV)$ and the far infrared colour indices are not completely independent entities
since $A(FUV)$ is measured using the total far infrared to UV flux ratio ($A(FUV)$ $\propto$ $L_{FIR}/L_{FUV}$), 
the strong correlations observed in the $S250/S500$ and $S250/S350$ vs. $A(FUV)$ cannot result from this underlying relation. 
This is due to the fact that i) the far infrared luminosity $L_{FIR}$ used to estimate $A(FUV)$ strongly depends on the total far infrared emission at $\lambda$ $\lesssim$ 100 $\mu$m, where most of the energy is radiated 
and only weakly on the dust emission at longer wavelengths, and ii) the SPIRE bands flux density ratio only depend at a second order on the 
total infrared luminosity that in our sample spans only three orders of magnitude. This last consideration is also valid for the $S60/S100$ colour index. 
The dependences of the IRAS/SPIRE flux density ratio on $A(FUV)$, on the contrary, might partly result from the fact that only IRAS flux densities are strongly related to the total
far infrared luminosity of galaxies. The weak dependence with metallicity observed in Fig. \ref{histo250500X} suggests that the far infrared colours of galaxies might also depend on 
the chemical properties of the dust grains.
\\
Overall, this analysis indicates that the emission properties of the cold dust dominating the far infrared spectral domain are 
regulated by the properties of the interstellar radiation field. Consistent with that observed in nearby, resolved galaxies (Bendo et al. 2011; Boquien et al. 2011), our analysis has shown
that the ionising and the non-ionising stellar radiation, including that emitted by the most evolved, cold stars, both contribute to the heating of the cold dust component.
This work, however, indicates that the mean metallicity of the gaseous phase of the interstellar medium is another key parameter characterising the cold dust emission of 
normal, late-type galaxies.

The simple interpretation outlined in this section, however, although based on solid statistical observational results, remains in a speculative stage given the strong degeneracy in the different 
parameters responsible for the emission of dust (dust grain emissivity, temperature, composition), and could be confirmed only after an accurate comparison with model
predictions. It is however clear that this observational evidence will be a major constraint in the years to come for models of dust emission in galaxies. \\

Considering the grain emissivity parameter, we obtain consistent results ($\beta$ $\simeq$ 1.5) with those obtained using similar $Herschel$ data in other nearby galaxies such as the 
LMC (Gordon et al. 2010) and M33 (Kramer et al. 2010). Values of $\beta$=1.2, 1.5 and 1.8 have been determined by the Planck collaboration for the SMC, LMC and the Milky Way, respectively
(Planck Collaboration 2011c).
Our value is however slightly larger than the one obtained by Planck for external galaxies once their 
SED are fitted with a single modified black body ($\beta$=1.2, $T$=26.3 K; Planck Collaboration 2011b).  
All these values of $\beta$ $\simeq$ 1.5 are significantly smaller than those obtained including far infrared data at shorter wavelengths ($\lambda$ $\leq$ 250 $\mu$m)
on a similar set of galaxies ($\beta$=2) by Davies et al. (2011; see also Fig. \ref{BB}). A decrease of $\beta$ with $\lambda$ has been previously reported by Paradis et al. (2009). 
They are, however, perfectly consistent with the values obtained by Dale et al. (2012) to fit the 100-500 $\mu$m SED of KINGFISH galaxies.

\section{Conclusions}

We have studied for the first time the far infrared (60-500 $\mu$m) colours of normal, late-type galaxies using new data recently obtained by the $Herschel$ space mission. 
We have determined 
the relationships between different far infrared colour indices, defined as flux density ratios using the two IRAS bands at 60 and 100 $\mu$m and the three SPIRE bands at 250, 350 and 500
$\mu$m, and compared them to various tracers of the physical properties of the target galaxies. These are
the present day star formation activity $SFR$, the birthrate parameter $b$ or equivalently the specific star formation rate $SSFR$, 
the surface brightness of the bulk of the stellar emission $\mu_e(H)$ and that of the ionising stellar radiation $\Sigma(H\alpha)$, the metallicity 12+log(O/H), and the internal attenuation $A(H\alpha)$ and $A(FUV)$.
Our analysis has shown that:\\
a) These far infrared colour indices are correlated with each other. Tight correlations are observed only for infrared indices close in $\lambda$. \\
b) The far infrared colour indices
are only partly related to the different tracers of the physical properties of the analysed galaxies. A tight correlation is observed only between the cold dust sensitive flux density ratios
$S250/S350$ and $S250/S500$ and the surface brightness of both the ionising and non-ionising radiation and $A(FUV)$. More dispersed relations are observed between the flux density ratios sensitive
to the position of the peak of the far infrared emission ($S60/S250$, $S100/S250$, $S100/S500$) and $\Sigma(H\alpha)$ and $\mu_e(H)$. Coarse relations are observed between the cold dust colour 
indices $S250/S350$ and $S250/S500$ and the metallicity 12+log(O/H), while the relationships between the other far infrared colours and the other physical tracers, if present, are very
dispersed.\\
c) We have shown that galaxies can be well segregated in the direction perpendicular to the main colour-colour relations 
determined using far infrared bands ($\lambda$ $\geq$ 100 $\mu$m) by means of their physical parameters, 
whose relative importance changes with the sampled spectral domain. Among these, the metallicity and the history of star formation, or equivalently the hardness of the interstellar
radiation field, seem the most important parameters.
The variation along the observed colour-colour relations, on the contrary, depends only weakly on all the physical parameters. \\
d) We have also shown that a single modified black body with a grain
emissivity index $\beta$=1.5 better fits the observed SPIRE flux density ratios $S250/S350$ vs. $S350/S500$ than $\beta$=2. Values of $\beta$ close to 2 are possible only in metal rich,
high surface brightness galaxies, while $\beta$ $\lesssim$ 1.5 are more representative of metal poor, low surface brightness objects.\\

These results are strong constraints for the study of the physical properties of the dust grains emitting in the far infrared spectral domain.
They first show that the properties of the large dust grains responsible for the emission in this spectral domain are not universal but rather change 
according to the physical properties of galaxies. The metallicity, the intensity of the ionising and non-ionising radiation fields as well as the hardness of the incident radiation
are indeed key parameters in regulating the dust emission in the far infrared spectral domain.

The observational evidence presented in this work is statistically significant. Its interpretation, however, here done only on phenomenological bases, 
requires an accurate comparison with different dust models for a more complete understanding of the properties of the cold dust dominating the emission of galaxies 
in the far infrared spectral domain. To this purpose, we have collected Spitzer data at shorter wavelengths 
(3 $\leq$ $\lambda$ $\leq$ 160 $\mu$m) to map the dust emission on the widest possible spectral range. These data, that will be soon combined with new PACS 
observations (Davies et al. 2011), will allow us to reconstruct the UV to radio continuum spectral energy distribution of this complete sample of galaxies necessary for a 
combined study of the heating sources and of the emitting dust (Ciesla et al., Boquien et al. in preparation). The comparison of the observed spectral energy distributions 
with different models available in the literature will definitely help us in making a further step in the comprehension of the emitting properties of 
the interstellar dust and their inter-relationships with the other physical parameters characterising the properties of the interstellar medium.  
The first observational evidence presented in this work might have strong implications in the study of galaxies at high redshift. They indeed show 
that the far infrared colours of galaxies, and thus their spectral energy distributions, might change as a function of the intensity of the interstellar 
radiation field and the metallicity, physical parameters that strongly evolve during the cosmic time. If possible variations in the star formation activity, 
and indirectly of the stellar radiation field, are often considered in cosmological studies (e.g Elbaz et al. 2011), metallicity variations are still neglected.
At the same time our results indicate that the assumption of a single modified black body with a grain emissivity parameter $\beta$=1.5 used to charaterise the far infrared-submillimetre
emission of galaxies at high redshift (e.g. Magdis et al. 2010a, 2010b; Hwang et al. 2010) is justified by the properties of nearby objects.

\begin{acknowledgements}

We thank the referee, D. Dale, for precious comments which helped improving the quality of the manuscript.
SPIRE has been developed by a consortium of institutes led
by Cardiff Univ. (UK) and including Univ. Lethbridge (Canada);
NAOC (China); CEA, LAM (France); IFSI, Univ. Padua (Italy);
IAC (Spain); Stockholm Observatory (Sweden); Imperial College
London, RAL, UCL-MSSL, UKATC, Univ. Sussex (UK); Caltech,
JPL, NHSC, Univ. Colorado (USA). This development has been
supported by national funding agencies: CSA (Canada); NAOC
(China); CEA, CNES, CNRS (France); ASI (Italy); MCINN (Spain);
SNSB (Sweden); STFC, UKSA (UK); and NASA (USA).
This research has made use of the NASA/IPAC Extragalactic Database 
(NED) which is operated by the Jet Propulsion Laboratory, California 
Institute of Technology, under contract with the National 
Aeronautics and Space Administration. The research leading to these results 
has received funding from the European Community's Seventh Framework Programme (/FP7/2007-2013/) under grant agreement No 229517.
This research has made use of the NASA/IPAC Extragalactic Database (NED) 
which is operated by the Jet Propulsion Laboratory, California Institute of 
Technology, under contract with the National Aeronautics and Space Administration
and of the GOLDMine database (http://goldmine.mib.infn.it/).

\end{acknowledgements}

\begin{appendix}
\section{Mutual correlations among the different physical parameters}

\begin{figure*}
   \centering
   \includegraphics[width=16cm]{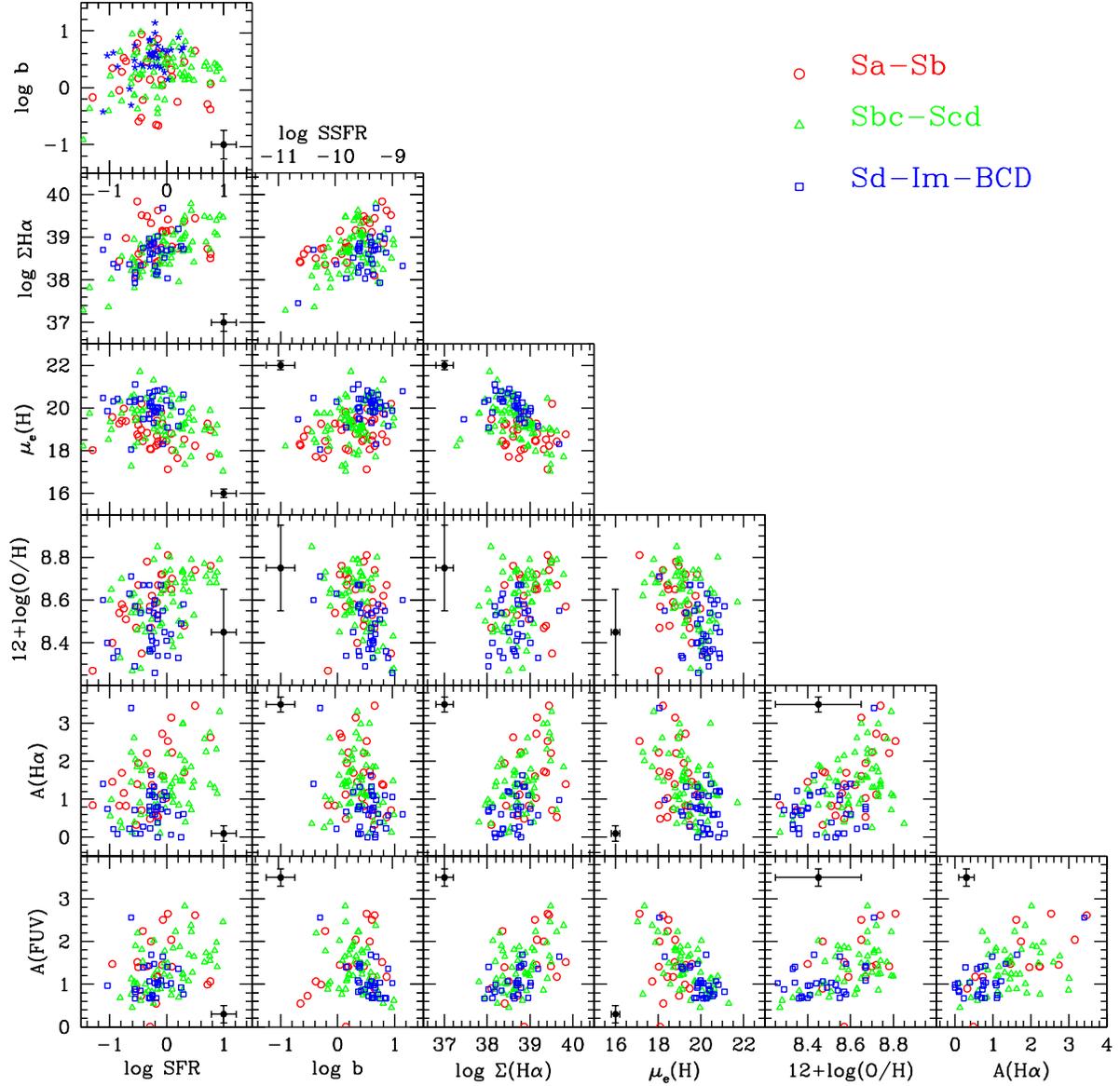}
   \caption{The relationships between the different parameters used to trace the physical properties characterizing the target galaxies: the logarithm of the star formation rate 
   $SFR$ (in M$_{\odot}$yr$^{-1}$), the logarithm of the 
   birthrate parameter $b$ (or $SSFR$), the logarithm of the H$\alpha$ effective surface brightness (in erg s$^{-1}$ kpc$^{-2}$), the H-band effective surface
   brightness (in AB mag arcsec$^{-2}$), the metallicity index 12+log(O/H), the Balmer decrement $A(H\alpha)$ (in magnitudes) and the FUV attenuation $A(FUV)$ (in magnitudes).
   Red open circles for Sa-Sb, green empty triangles for Sbc-Scd and blue open squares for Sd-Im-BCD. The typical error bar is indicated with a black cross.
   }
   \label{fisicicor}%
   \end{figure*}

The physical parameters used to trace the properties of the galaxies analysed in this work, although determined using independent data, can trace 
non completely independent variables. The $SFR$, the birthrate parameter $b$ and the surface brightness of the ionising radiation $\Sigma(H\alpha$), for instance, are 
all strongly related to the present day star formation activity. The effective surface brightness $\mu_e(H)$ traces the distribution of the evolved stars
dominating the stellar mass of galaxies. The total stellar mass is also necessary to estimate the birthrate parameter. $A(H\alpha)$ and $A(FUV)$
are two independent tracer of the dust extinction within galaxies. These different physical parameters might thus be mutually related, 
as shown in Fig. \ref{fisicicor} and Table \ref{Tabfisicicor}. \\
Figure \ref{fisicicor} shows a tight correlation between the surface brightness of the very evolved, cold stars $\mu_e(H)$ and the UV attenuation $A(FUV)$ ($r$ = -0.57)
indicating that in normal, star forming galaxies dust absorbs at the same time the photons emitted by the newly formed stars and by the most evolved stellar populations. 
$\Sigma(H\alpha$), $SFR$ and $b$, as previously mentioned, are not fully independent tracers since all related with the present day star formation activity of galaxies.
Surprisingly, however, the mutual relations between these parameters are very dispersed.
The correlation between $\mu_e(H)$ and $\Sigma(H\alpha$) suggests that the surface density of all kind of stars 
increases independently of their age in the galaxies analysed in this work.
Naturally, the two independent dust extinction tracers $A(H\alpha)$ and $A(FUV)$ are mutually related (e.g. Calzetti 2001; Boselli et al. 2009).
The trends observed between $A(FUV)$ (and $A(H\alpha)$) and $\Sigma(H\alpha$) or $\mu_e(H)$ suggest that all the stellar radiation participates to the heating of the dust
emitting in the far infrared. The relations with the metallicity index 12+log(O/H) are weak ($r$ $\simeq$ 0.4-0.5), and suggest, as expected, that the dust extinction is more important
in metal rich objects. They also show that the metallicity is coupled with the star formation history of galaxies ($b$), and is generally higher in high surface brightness 
evolved systems than in star forming, low surface brightness objects.

\begin{table*}
\caption{Spearman correlation coefficients of the relations between the different physical parameters (Fig. \ref{fisicicor})}
\label{Tabfisicicor}
{
\[
\begin{tabular}{ccccccc}
\hline
\noalign{\smallskip}
Y-colour		& log $SFR$		& log $b$	& log $\Sigma(H\alpha)$			& $\mu_e(H)$		& 12+log(O/H)	& $A(H\alpha)$  \\
Units			& M$_{\odot}$ yr$^{-1}$	&		& erg s$^{-1}$ kpc$^{-2}$		& AB mag arcsec$^{-2}$	&		& mag	          \\
\hline
log $b$			& 0.11			& 		& 					& 			& 	   	&    \\
log $\Sigma(H\alpha)$	& 0.56			& 0.24		& 					& 			& 	   	&    \\
$\mu_e(H)$		& -0.25			& 0.35		& -0.39					& 			&  		&    \\
12+log(O/H)		& 0.44			& -0.55		& 0.43					& -0.50			&  		&    \\
$A(H\alpha)$		& 0.34			& -0.28		& 0.33					& -0.43			& 0.47		&    \\
$A(FUV)$		& 0.36			& -0.38		& 0.45					& -0.57			& 0.54		& 0.56	       \\
\noalign{\smallskip}
\hline
\end{tabular}
\]
}
\end{table*}

\section{Colour-colour diagrams vs. physical parameters}

\begin{figure*}
   \centering
   \includegraphics[width=16cm]{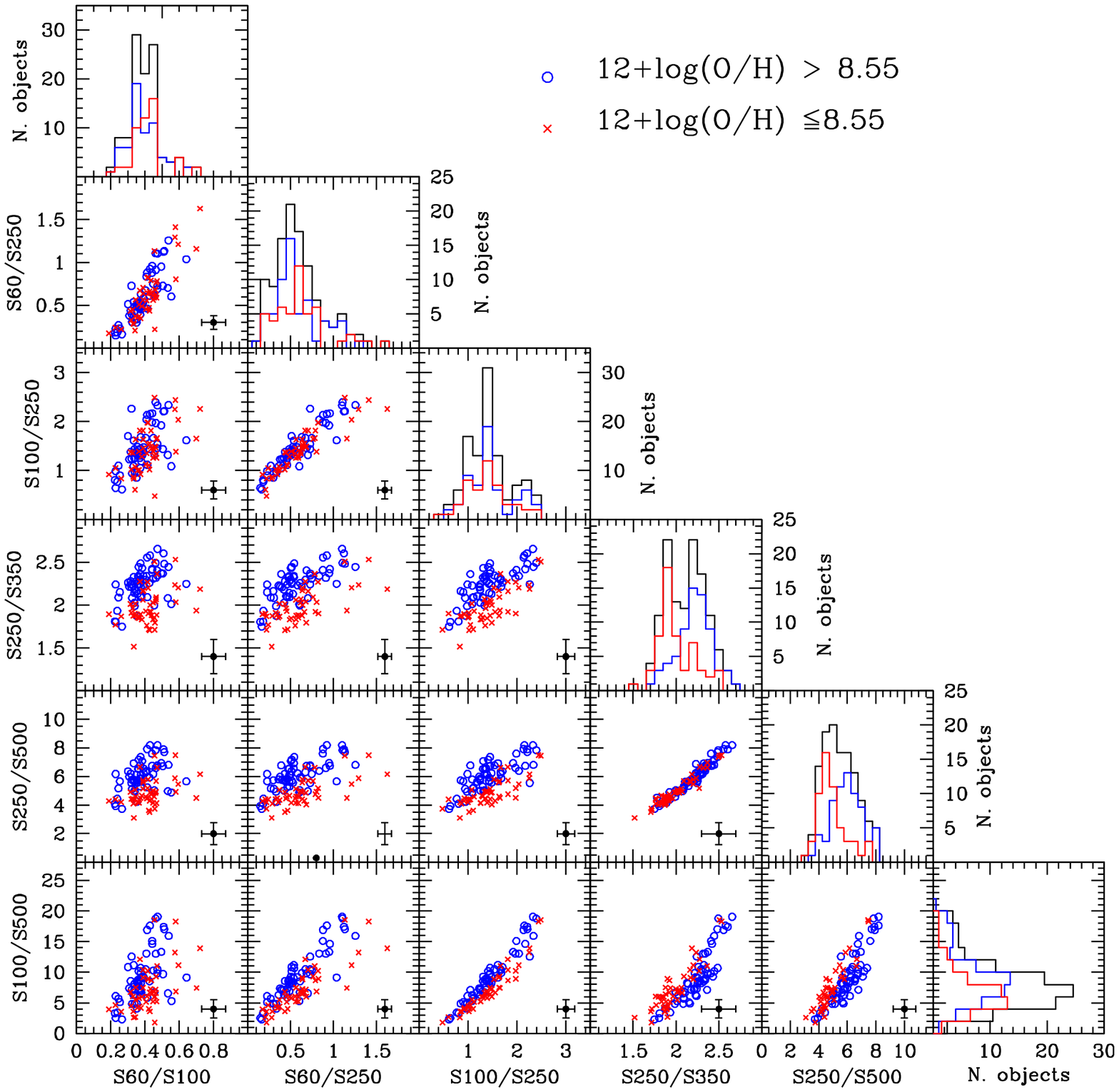}
   \caption{Far infrared colour-colour diagrams, equivalent to those shown in Fig. \ref{colours}, with 
   galaxies coded according to their mean metal content, with blue open circles for metal rich galaxies (12+log(O/H)$>$ 8.55) and red crosses for metal poor objects (12+log(O/H)$\leq$
   8.55). The black histogram gives the distribution of all galaxies along the X-axis, while the coloured histograms those of the two subsamples of galaxies selected according to their metallicity.
   }
   \label{metal}%
   \end{figure*}

\begin{figure*}
   \centering
   \includegraphics[width=16cm]{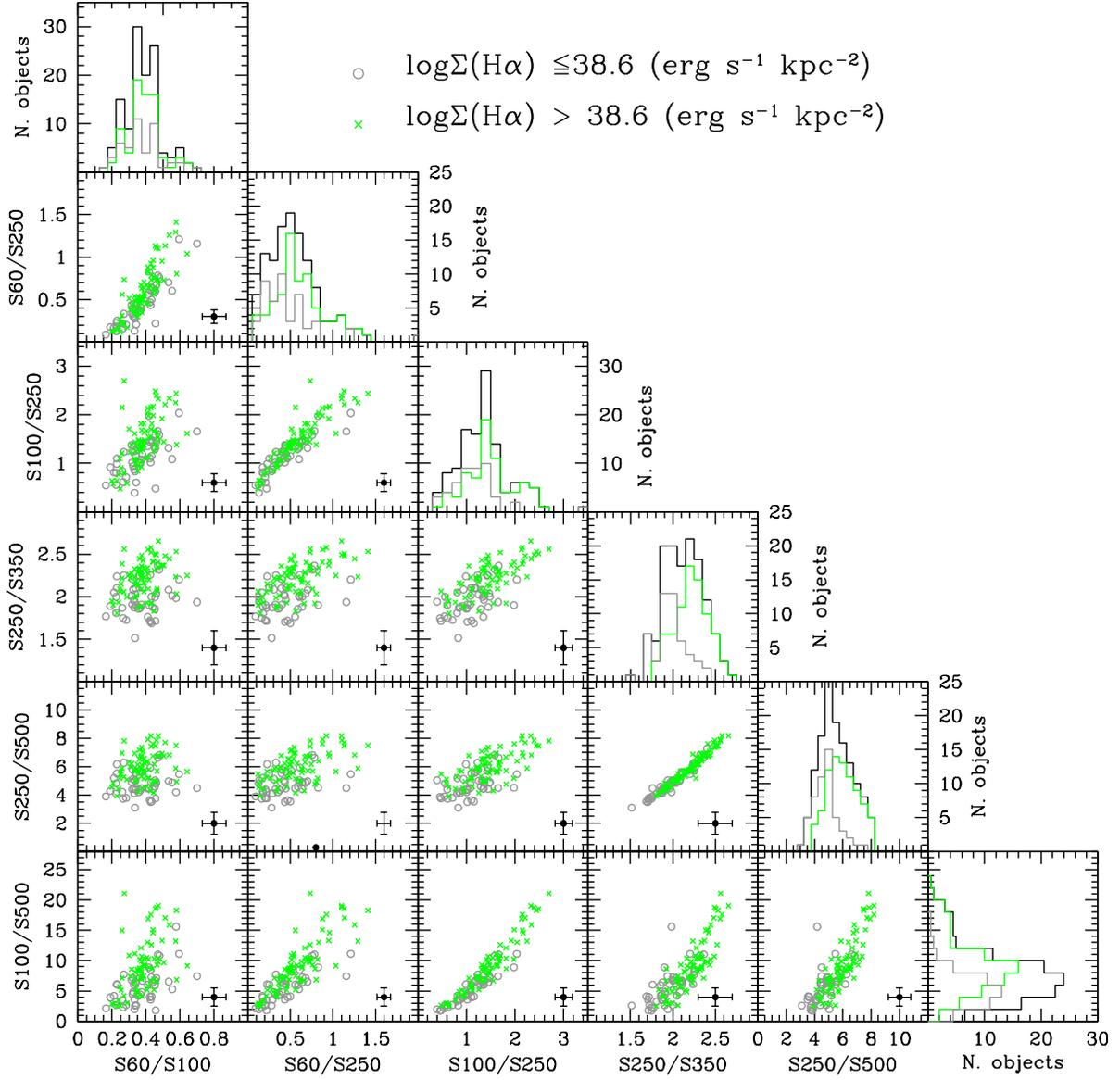}
   \caption{Far infrared colour-colour diagrams, equivalent to those shown in Fig. \ref{colours}, with 
   galaxies coded according to their H$\alpha$ surface brightness, with green crosses for objects with log $\Sigma(H\alpha)$ $>$ 38.6 erg s$^{-1}$ kpc$^{-2}$
   and grey open circles for galaxies with log $\Sigma(H\alpha)$ $\leq$ 38.6 erg s$^{-1}$ kpc$^{-2}$. The black histogram gives the  
   distribution of all galaxies along the X-axis, while the coloured histograms those of the two subsamples of galaxies selected according to their
   H$\alpha$ surface brightness.}
   \label{SHa}%
   \end{figure*}

\begin{figure*}
   \centering
   \includegraphics[width=16cm]{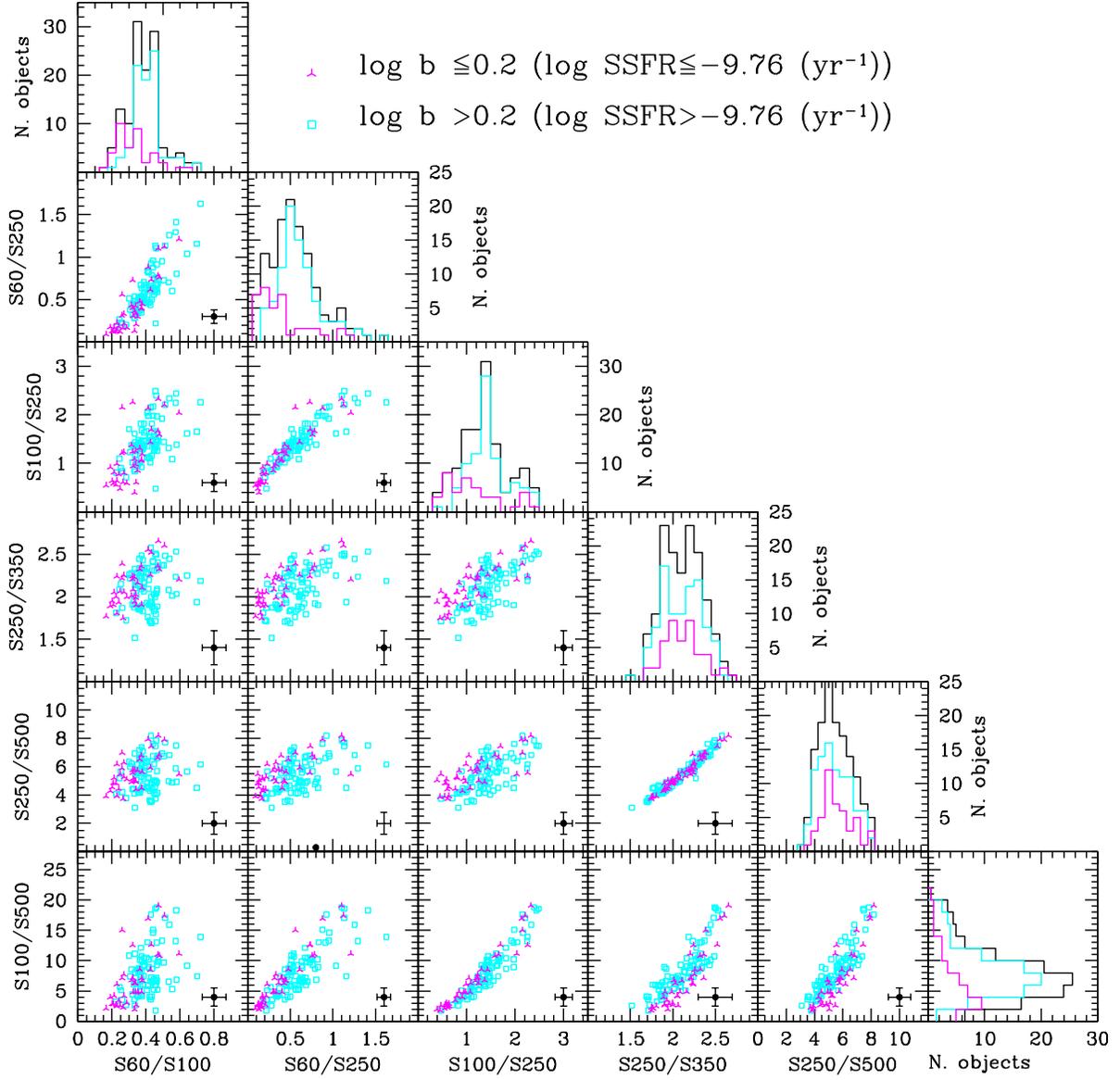}
   \caption{Far infrared colour-colour diagrams, equivalent to those shown in Fig. \ref{colours}, with 
   galaxies coded according to their birthrate parameter, with cyan open squares for objects with log $b$ $>$ 0.2 and magenta three points stars for galaxies with log $b$ $\leq$ 0.2.
   The black histogram gives the distribution of all galaxies along the X-axis, while the coloured histograms those of the two subsamples of galaxies selected according to their
   birthrate parameter.  
   }
   \label{bpar}%
   \end{figure*}

\begin{figure*}
   \centering
   \includegraphics[width=16cm]{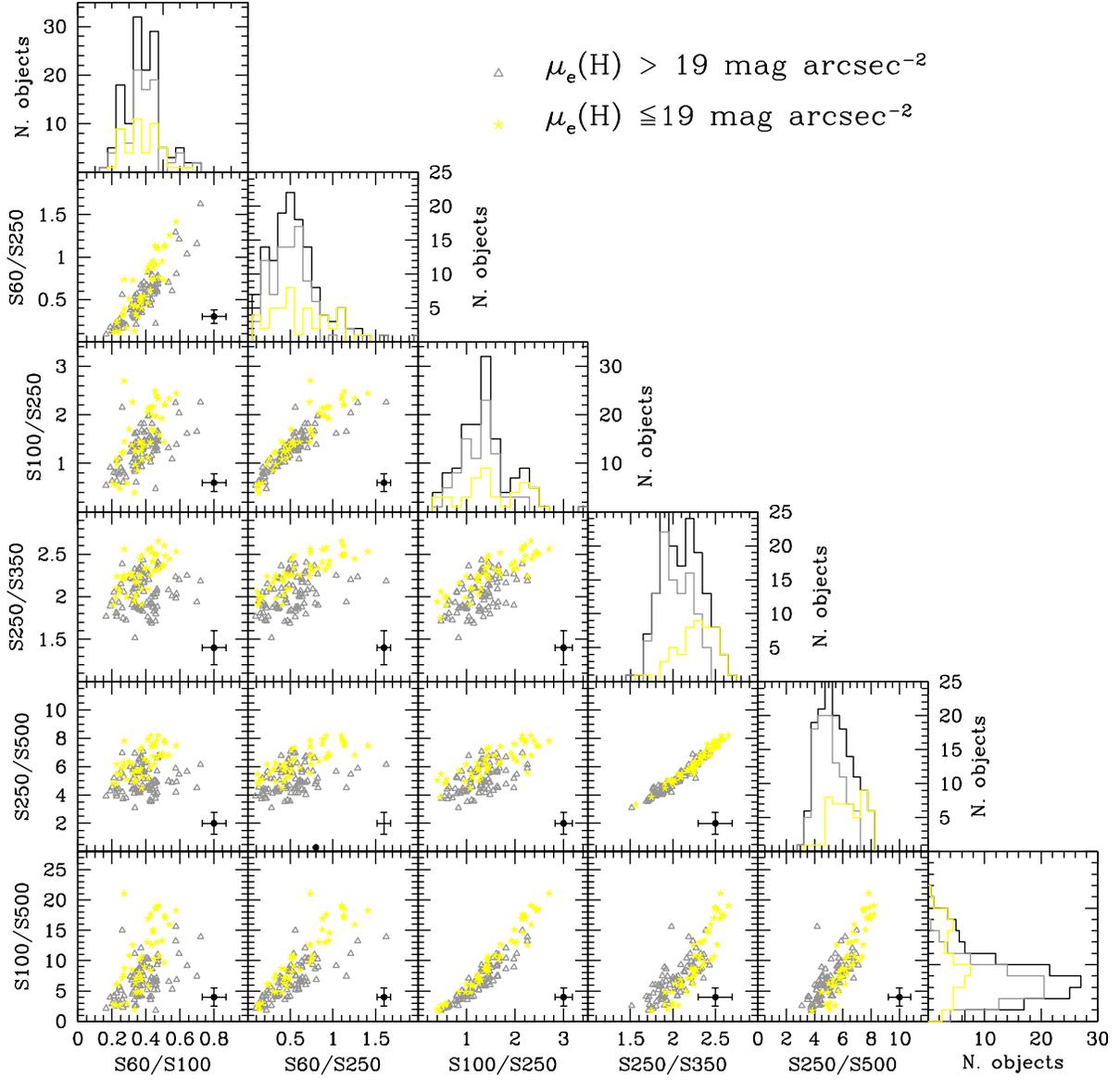}
   \caption{Far infrared colour-colour diagrams, equivalent to those shown in Fig. \ref{colours}, with 
   galaxies coded according to their effective surface brightness, with grey open triangles for objects with $\mu_e(H)>$ 19 mag arcsec$^{-2}$ and
   yellow asterisks for galaxies with $\mu_e(H) \leq$ 19 mag arcsec$^{-2}$. The black histogram gives the 
   distribution of all galaxies along the X-axis, while the coloured histograms those of the two subsamples of galaxies selected according to their
   H band effective surface brightness.
   }
   \label{mueh}%
   \end{figure*}

\begin{figure*}
   \centering
   \includegraphics[width=16cm]{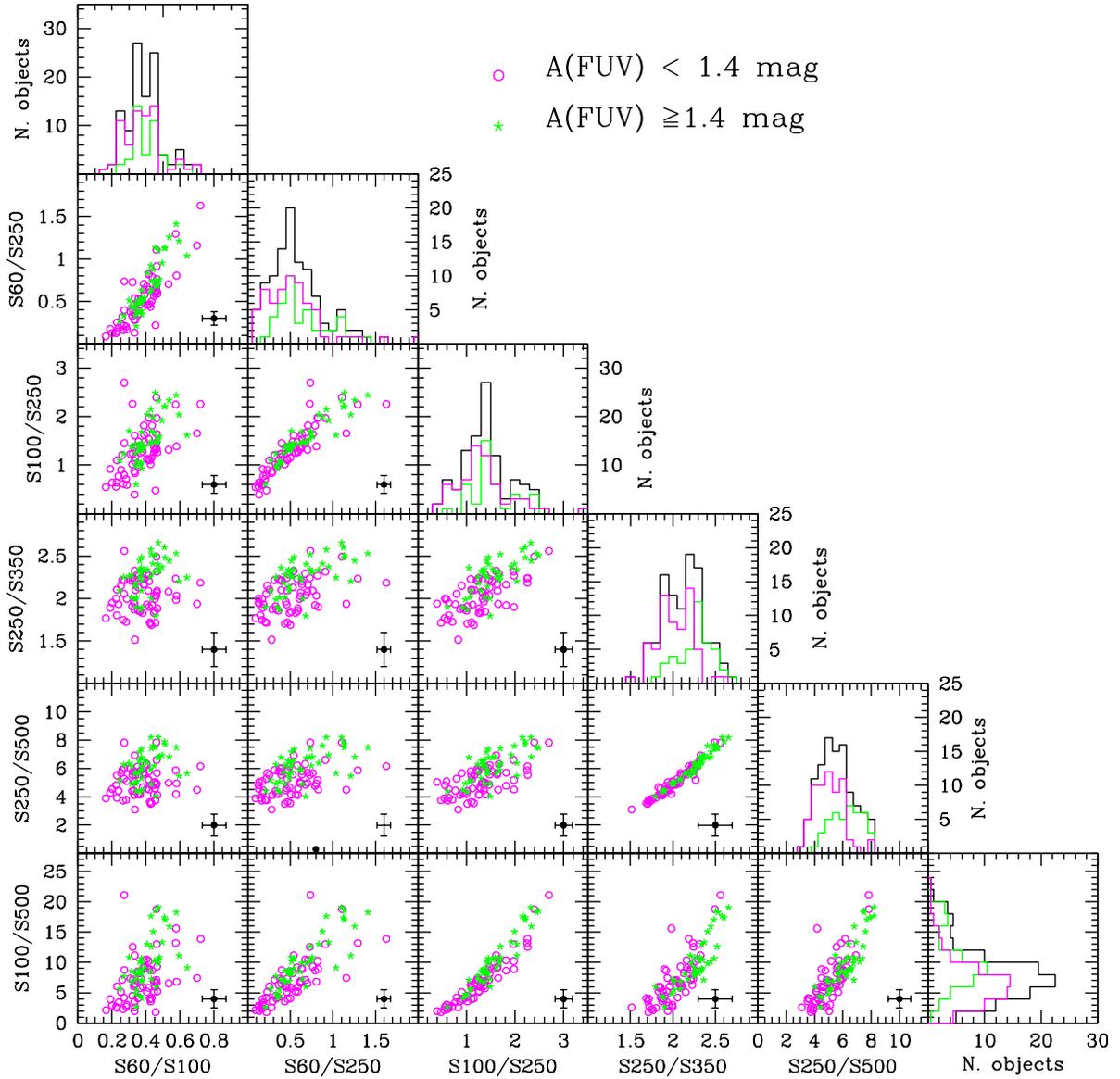}
   \caption{Far infrared colour-colour diagrams, equivalent to those shown in Fig. \ref{colours}, with 
   galaxies coded according to their FUV attenuation, magenta open circles for galaxies with a low attenuation ($A(FUV)<$1.4) and green asterisks for objects wit $A(FUV)\geq$ 1.4.
   The black histogram gives the distribution of all galaxies along the X-axis, while the coloured histograms those of the two subsamples of galaxies selected according to their FUV attenuation.}
   \label{AFUV}%
   \end{figure*}

To extend the analysis done in section 4, we plot here for completeness all the colour-colour diagrams shown in Fig. \ref{colours}
for galaxies coded according to different physical parameters. The codes used are the same than those used in Fig. 
\ref{histo60100X}, \ref{histo250500X}, \ref{histo100250Y} and \ref{histo250500Y}, where the threshold in the physical parameters are: 
metal content 12+log(O/H) = 8.55; H$\alpha$ surface brightness log $\Sigma(H\alpha$) = 38.6 erg s$^{-1}$ kpc$^{-2}$; birthrate parameter log $b$ = 0.2; 
H-band effective surface brightness $\mu_e(H)$ = 19 AB mag arcsec$^{-2}$; FUV attenuation $A(FUV)$ = 1.4 mag. The general behaviors described in section 4 can be observed also 
in the Fig. \ref{metal}, \ref{SHa}, \ref{bpar}, \ref{mueh}  and \ref{AFUV}.

\end{appendix}

\end{document}